\documentclass[12pt]{article} 

\usepackage[utf8]{inputenc} 

\RequirePackage[OT1]{fontenc}
\RequirePackage{amsthm,amsmath,amssymb,amstext,graphicx,psfrag}
\RequirePackage[colorlinks]{hyperref}
\usepackage{turnstile}
\usepackage{graphicx}
\usepackage{amssymb,amsmath,amsfonts}
\usepackage{enumerate}

\usepackage{setspace}
\usepackage{titling}
\usepackage{graphicx} 

\usepackage{booktabs} 
\usepackage{array} 
\usepackage{paralist} 
\usepackage{verbatim} 
\usepackage{subfig} 

\usepackage{fancyhdr} 
\usepackage{sectsty}
\allsectionsfont{\sffamily\mdseries\upshape} 

\usepackage[nottoc,notlof,notlot]{tocbibind} 
\usepackage[titles,subfigure]{tocloft} 

\usepackage{amsmath}
\usepackage{amssymb}
\usepackage{amsthm}
\usepackage{rotating}
\usepackage{textcomp}
\usepackage{algorithmic}
\usepackage{algorithm}
\newtheorem{1}{Lemma}

\usepackage{epsfig}
\usepackage{graphics}
\usepackage[tmargin=1in, bmargin=1in, rmargin=1in, lmargin=1in]{geometry}
\usepackage{setspace}
\usepackage{fancyhdr}
\usepackage[dvipsnames,usenames]{color}

\usepackage[authoryear]{natbib}
\bibpunct{(}{)}{;}{a}{}{,}

\newcommand{\iid}{\stackrel{\emph{i.i.d.}}{\sim}}

\numberwithin{equation}{section}
\theoremstyle{plain}

\title{ \Large{ A Methodology for Robust Multiproxy Paleoclimate Reconstructions \& Modeling of Temperature Conditional Quantiles }
}
\author{Lucas Janson and Bala Rajaratnam\\
Stanford University
}

\begin{document}

\vspace{-5cm}

\maketitle

\thispagestyle{empty}



\onehalfspacing

\begin{abstract}
Great strides have been made in the field of reconstructing past temperatures based on models relating temperature to temperature-sensitive paleoclimate proxies.  One of the goals of such reconstructions is to assess if current climate is anomalous in a millennial context. These regression based approaches model the conditional mean of the temperature distribution as a function of paleoclimate proxies (or vice versa). Some of the recent focus in the area has considered methods which help reduce the uncertainty inherent in such statistical paleoclimate reconstructions, with the ultimate goal of improving the confidence that can be attached to such endeavors. A second important scientific focus in the subject area is the area of forward models for proxies, the goal of which is to understand the way paleoclimate proxies are driven by temperature and other environmental variables.  One of the primary contributions of this paper is novel statistical methodology for (1) quantile regression with autoregressive residual structure, (2) estimation of corresponding model parameters, (3) development of a rigorous framework for specifying uncertainty estimates of quantities of interest, yielding (4) statistical byproducts that address the two scientific foci discussed above. We show that by using the above statistical methodology we can demonstrably produce a more robust reconstruction than is possible by using conditional-mean-fitting methods. Our reconstruction shares some of the common features of past reconstructions, but we also gain useful insights. More importantly, we are able to demonstrate a significantly smaller uncertainty than that from previous regression methods. In addition, the quantile regression component allows us to model, in a more complete and flexible way than least squares, the conditional distribution of temperature given proxies. This relationship can be used to inform forward models relating how proxies are driven by temperature.
\end{abstract}

{\bf Keywords}: quantile regression with time series errors, principal components, uncertainty quantification, statistical paleoclimate reconstructions.
\thispagestyle{empty}
\newpage
\pagenumbering{arabic}

\section{Introduction}
The study of climate over the Earth's history is a topic of interest whose relevance has increased rapidly with the growing concern over anthropogenic global warming (AGW), or more formally known in the scientific literature, as climate change due to anthropogenic causes (CC).  Within the wide realm of climate, the study of Earth's \emph{temperature} is of central importance, as even small changes in temperature over an extended period of time can have drastic effects on ecosystems (loss of suitable habitat) \citep{Mann1998}, agriculture (crop yields) \citep{Porter2005, Sheehy2006, Schlenker2009, Solomon2007}, and even geography (melting ice caps, rising water levels) \citep{Mann1998}.

An important component in the study of Earth's temperature is the reconstruction of past temperatures using paleoclimate proxies.  Paleoclimate proxies are essentially natural record-keepers that are understood to be related to temperature, can be resolved at some known temporal resolution, and can reach back centuries.  For example, tree-ring widths are known to be related to temperature, and are distinguishable at an annual resolution. Hence by taking a core from the trunk of a very old tree, insight into historical temperatures can be gained.  Other proxies include marine sediments, ice cores, and coral.  By calibrating a statistical model between the proxies and known temperatures over the period where there exists a reliable temperature record, reconstructions of past climate can be made reaching back as far as the proxies do, with some going back more than a millennium.

For better or for worse, the results of past temperature reconstructions have been used to inform policy.  Although General Circulation Models / Global Climate Models (GCMs) consider the case for AGW in the future, reconstructions of past temperatures, such as that in \citep{Mann1998}, can provide jarring and easily interpretable suggestion from a different perspective.  In the interest of properly informing policy-makers and the public that elects them, it is crucial that such reconstructions be done with the utmost statistical rigor.  In addition, reconstructions of past climate can help validate the GCMs that predict future temperatures. As it were, the statistical multiproxy paleoclimate reconstruction problem is a very challenging one, as there is relatively little data, high noise, and significant spatial and temporal dependence to be accounted for. 

Many authors have made huge strides in the task of modeling mean annual temperature over approximately the past millennium and beyond.  We avoid an extensive literature review here, and refer the reader to the very useful work of \cite{Tingley2012} for a comprehensive overview.  The exact nature of the hindcasts of temperature time series varies from author to author, with some focusing on just global or hemispherical means, see references 3-14 of \cite{Mann2008}, while others approach the problem of predicting regional mean annual temperatures across the globe \citep{Mann1998, Mann2008, Rutherford2005}. A few authors have even attempted the difficult task of reconstructing climates of the past over the entire globe using paleoclimate proxies \citep{Tingley2010, Tingley2010a}.  Although these various approaches differ in their modelling assumptions and specific statistical methodologies, to our knowledge they all attempt to fit the conditional mean of the temperature using some variant of least squares regression (LS).  A related question in this regard also pertains to investigating methods that reduce the uncertainty inherent in such reconstructions.

One of the key features of this paper is the development of a novel statistical methodology which combines quantile regression with autoregressive (AR) structure in the residuals. In addition, an algorithm for estimation of model parameters is proposed and analyzed. In order to undertake uncertainty quantification, a comprehensive framework for statistical inference of these parameters is developed and analyzed.  Although the key contribution of the paper is the proposed statistical methodology, the motivation behind the development of the method is the important application for which it is well-suited.

This paper seeks to approach reconstructing Northern Hemisphere (NH) temperatures over the past millenium through temperature-related proxies by a more robust approach than has been taken in previous work.  This is achieved through our time-series version of quantile regression (QR), with two useful end products.  We show that in our millenial reconstruction, there is a significant statistical payoff from the robustness of quantile regression as compared to least squares techniques.  In addition, by using the fact that quantile regression can be applied to any quantile from 0 to 1, we can model the entire conditional distribution of temperature given the proxies, instead of being restricted to a parameterized distribution.

The stochastic modeling of how temperature sensitive climate proxies are causally affected by climate processes is also an important area of research within the broader statistical paleoclimate community, see \cite{Tingley2012} for an instructive overview. From an earth sciences perspective the need to model and understand how a given proxy varies as a function of temperature and climate is compelling, and is a complex, intricate and only partially understood process. Such models are often  referred to as \emph{forward models} in the literature and are used to specify the mechanisms which generate data in the paleoclimate context. Constructing forward models requires significant scientific knowledge, as each proxy type responds differently to environmental factors. Hence proxy-specific forward models are required in order to gain meaningful insight into them. Clearly, constructing forward models endeavors have tremendous values for both understanding how each proxy type record temperature and also their ability to do so.  Several authors have looked into the specification of functional forms for forward models. Both  linear and non-linear models have been considered in the literature, see \cite{Evans2006, Tolwinski-Ward2011} to name just a few. These functional forms aim to describe the conditional distribution of the proxies given the temperature, i.e., $f(P|T)$. Modeling this conditional distribution in a flexible way that incorporates differences at the extreme quantiles has the ability to explain how different proxies respond to extreme temperatures. We shall demonstrate that the methodology developed in this paper is useful for obtaining conditional quantile estimates.

The remainder of the paper is organized as follows. Section 2 introduces the quantile regression approach with autoregressive structure in the residuals. Estimating model parameters and undertaking uncertainty quantification of parameter estimates are also investigated in detail. Section 3 introduces the climate data that is used in our application, as well as some data-specific considerations for applying the above methodology.  Section 4 presents a millenial reconstruction, and demonstrates the improvement over analogous least squares techniques.  Section 5 shows the result of modelling a range of quantiles of the conditional distribution of temperature given the proxies.  Section 6 discusses conclusions and possible future directions.

\section{Methodology}
\subsection{Theory of Quantile Regression}
Readers already familiar with quantile regression can skip this subsection. While least squares regression (LS) fits a model for the conditional mean of a response variable given some predictor variables, quantile regression (QR) fits a model for some conditional quantile $\tau \in (0,1)$ of a response variable given some predictor variables.  The two are similar in that both solve a minimization problem where the objective function is the sum of some norm of the residuals of the model.  The method of least squares minimizes the sum of squared residuals, while QR, in the $\tau=0.5$ case, minimizes the sum of the absolute values of residuals and corresponds to the method of least absolute deviations.  In the general $\tau$ case, the absolute value function $|x|$ is replaced by the so-called `check' function \citep{Koenker2005}, see Figure \ref{checkplot}:

\begin{figure}\centering
\includegraphics[scale=.4]{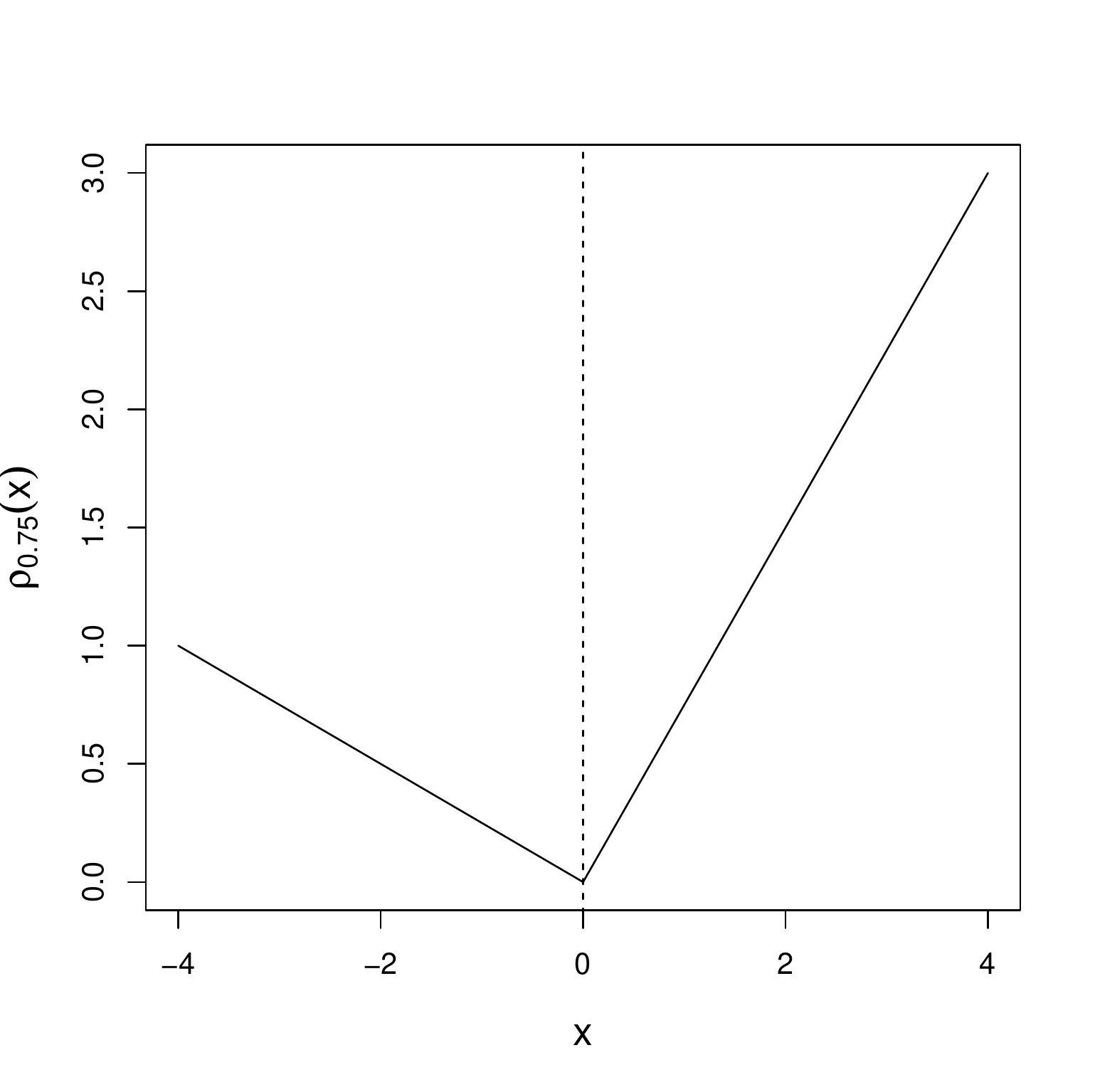}
\caption{An example of the check function for $\tau=0.75$.}
\label{checkplot}
\end{figure}

\begin{equation}
\label{check}
\rho_\tau(x)=\left\{\begin{array}{rl} \tau x & $if $x\ge 0, \\ (\tau-1)x & $if $x<0. \end{array} \right.
\end{equation}
Thus the QR estimate $\hat{\boldsymbol{\beta}}_{\tau}$ of the conditional $\tau^{th}$ quantile of a response $y$ given predictors $\boldsymbol{x}$ is
\begin{equation}
\label{QRopt}
\hat{\boldsymbol{\beta}}_{\tau} = \bold\arg\min_{\boldsymbol{\beta} \in \mathbb{R}^{p+1}} \sum_{i=1}^n \rho_\tau(y_i - \boldsymbol{x}_i^\top \boldsymbol{\beta}),
\end{equation}
where $n$ is the number of observations $(y_i, \boldsymbol{x}_i)$ and $p$ is the number of predictors. Note we have absorbed the intercept into $\boldsymbol{\beta}$ as its first component, and thus treat the $\boldsymbol{x}_i$ as $(p+1)$-vectors with first component equal to one.

From (\ref{QRopt}) it is evident that separate models can be fitted for each conditional quantile.  Contributions of each predictor can be analyzed at different quantiles in order to explore relationships beyond that with the conditional mean, the only relationship revealed by standard least squares regression.  Also note that the estimator in (\ref{QRopt}) is equivalent to the maximum likelihood estimator (MLE) for a linear model with \emph{i.i.d} residuals with probability density function:

\begin{equation}
\label{chexponential}
f_{\varepsilon}^{(\tau)} (z) =\frac{\tau(1-\tau)}{b} \cdot \exp \bigg( \frac{-\rho_\tau(z)}{b} \bigg),
\end{equation}
for some $b>0$. The residual distribution in (\ref{chexponential}) reduces to the Laplace distribution when $\tau=0.5$.  The above also reveals the robust nature of quantile regression, as this residual distribution has heavier tails than the Gaussian distribution.

One of the setbacks of QR as compared to LS is the computations required to obtain the quantile regression estimator, $\hat{\boldsymbol{\beta}}_{QR}$.  In particular, from basic linear algebra, LS has a closed form solution for the least squares estimator, $\hat{\boldsymbol{\beta}}_{LS}$, but QR is traditionally solved by computational optimization methods such as linear programming.  The reader is referred to \citep{Koenker1987} for insightful ways to exploit the structure of the QR problem to improve the efficiency of the straightforward linear programming approach.  

\subsection{Modelling Autoregressive Structure in Residuals}
\begin{figure}\centering
\includegraphics[scale=.4]{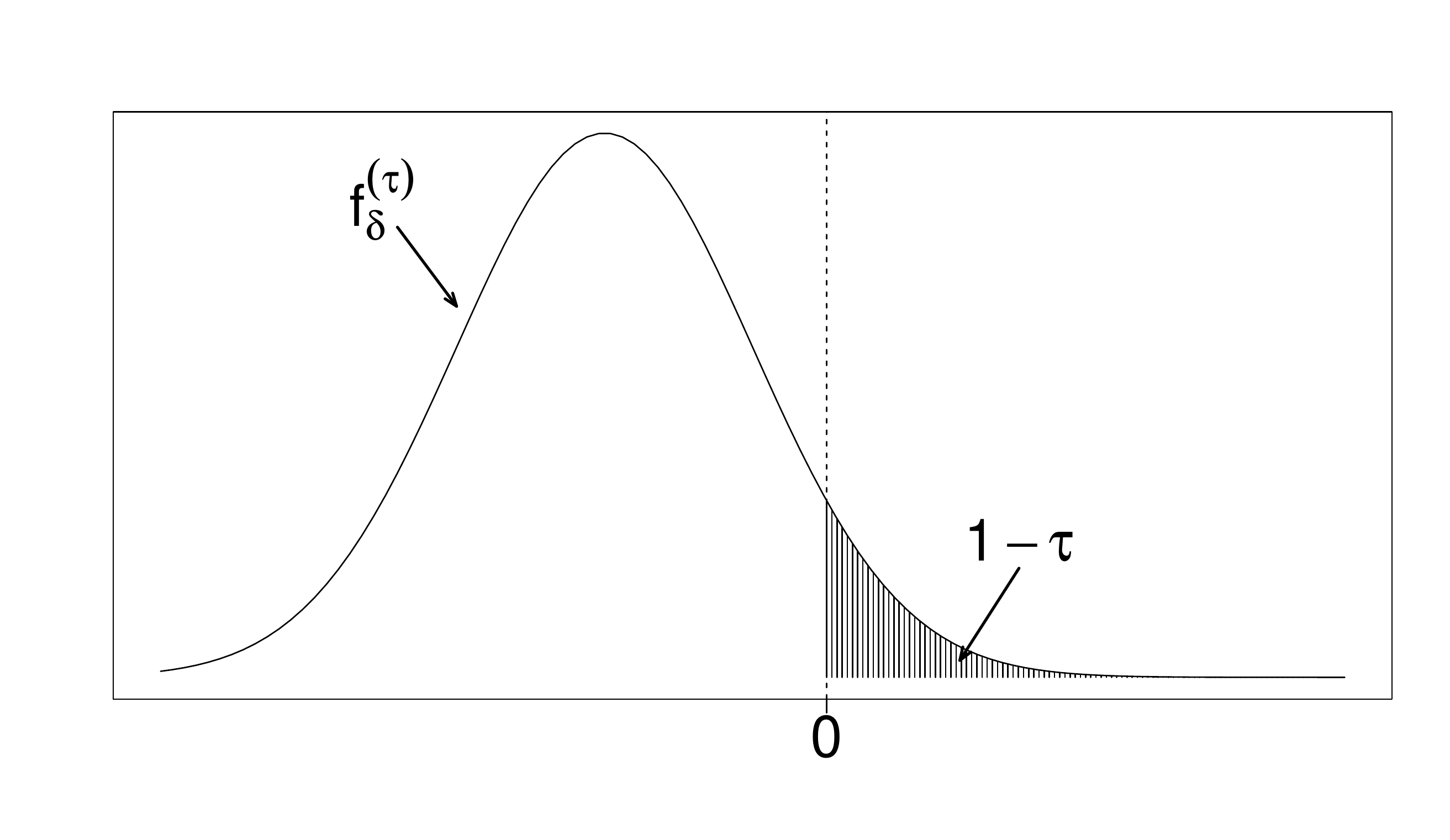}
\caption{Schematic for the distribution of the $\delta_i^{(\tau)}$.}
\label{zeroQuantilePlot}
\end{figure}

\cite{Li2007} note that the residuals of models fit to paleoclimate proxies are not independent, and in fact tend to exhibit autoregressive (AR) behavior.  Other authors have also found low order autoregressive structure in the residuals \citep{McShane2011, Tingley2010, Tingley2010a, Tingley2012}.  To this end, we propose the linear QR model with AR($q$) residuals for the $\tau^{th}$ conditional quantile of the response as follows:
\begin{equation}
\label{QR}
y_i = \boldsymbol{x}_i^\top \boldsymbol{\beta}^{(\tau)} + \varepsilon_i^{(\tau)}
\end{equation}
with $\varepsilon_i^{(\tau)}$ an AR($q$) process under the model
\begin{equation}
\label{AR}
\varepsilon_i^{(\tau)} = \phi_1^{(\tau)} \varepsilon_{i-1}^{(\tau)} + \cdots + \phi_q^{(\tau)} \varepsilon_{i-q}^{(\tau)} + \delta_i^{(\tau)}
\end{equation}
where the $\delta_i^{(\tau)}$ are \emph{i.i.d.} with some distribution $f_{\delta}^{(\tau)}$ whose $\tau^{th}$ quantile is zero, see Figure \ref{zeroQuantilePlot}.  Thus the conditional $\tau^{th}$ quantile of the response, $Q^{(\tau)}(y_i) | \boldsymbol{x}_i$ is given by
\begin{equation*}
Q^{(\tau)}(y_i) | \boldsymbol{x}_i = \boldsymbol{x}_i^\top \boldsymbol{\beta}^{(\tau)} + \phi_1^{(\tau)} \varepsilon_{i-1}^{(\tau)} + \cdots + \phi_q^{(\tau)} \varepsilon_{i-q}^{(\tau)}.
\end{equation*}
Note that through the AR structure in the residuals, we have also implicitly conditioned $Q^{(\tau)}(y_i) | \boldsymbol{x}_i$ on some previous data, namely, $y_{i-q}, \dots, y_{i-1}, \boldsymbol{x}_{i-q}, \dots, \boldsymbol{x}_{i-1}$. It is important to recognize explicitly that the conditional $\tau^{th}$ quantile is a parameter of the conditional distribution (as compared to a predicted value).

In recent work, \cite{Koenker2006} provide a useful method for fitting the quantiles of a univariate autoregressive time series.  In this paper however, we seek to address the different problem of relating predictors to a response variable. In particular, we provide a statistical methodology and an associated algorithm for multivariate QR with AR residual structure.  Reformulating the QR optimization problem to accommodate the AR residual structure as given in (\ref{AR}) unfortunately results in a nonconvex objective function.  This is formally stated in the following Lemma.  
\begin{1}
\label{lemma}
Consider the quantile regression model above in (\ref{QR}) and (\ref{AR}) with autoregressive residual structure of order $q \ge 1$.  Let $\{(y_1, \boldsymbol{x}_1), \dots, (y_n, \boldsymbol{x}_n)\}$, $\boldsymbol{x}_i \in \mathbb{R}^p$, $n>p+q$ be i.i.d. observations and let coefficient estimates $\hat{\boldsymbol{\phi}}_{\tau}, \hat{\boldsymbol{\beta}}_{\tau}$ be given by the following optimization:
\begin{align}
[\hat{\boldsymbol{\phi}}_{\tau}, \hat{\boldsymbol{\beta}}_{\tau}] & = \arg \min_{\boldsymbol{\phi} \in \mathbb{R}^q, \boldsymbol{\beta} \in \mathbb{R}^{p+1}} \sum_{i=q+1}^n \rho_{\tau} (\delta_i) \nonumber \\
& = \arg \min_{\boldsymbol{\phi} \in \mathbb{R}^q, \boldsymbol{\beta} \in \mathbb{R}^{p+1}} \sum_{i=q+1}^n \rho_{\tau} (\phi(B) y_i - \phi(B)\boldsymbol{x}_i^\top \boldsymbol{\beta}),
\label{ARQR}
\end{align}
where $\rho_{\tau}$ is the check function defined in Section 2.1, $B$ the backshift operator, and $\phi(.)$ the AR polynomial defined by $\phi(z) = 1- \phi_1 z - \cdots - \phi_q z^q$ with $\boldsymbol{\phi} = (\phi_1,\dots,\phi_q)$.  Assume that the predictors satisfy the mild regularity condition that there exists some $j \in \{1,\dots,p\}$ such that
\begin{equation}
\sum_{i = q+1}^n a_i x_{i,j} \ne 0 \quad \forall (a_{q+1},\dots,a_n) \in \{1-\tau,\tau\}^{n-q},
\label{ARQRassumpt}
\end{equation}
where $x_{i,j}$ refers to the $j^{th}$ component of the $i^{th}$ observation.  Then the objective function in (\ref{ARQR}) above is not convex.
\begin{proof}
See Appendix A.
\end{proof}
\end{1}
{\it Remark:} Note that for a given $\boldsymbol{x}_i$, condition (\ref{ARQRassumpt}) only excludes $2^{n-q}$ one-dimensional sets from $\mathbb{R}^{n-q}$. So if $\boldsymbol{x}_i$ is drawn from a continuous distribution on $\mathbb{R}^{n-q}$, this constraint will be satisfied with probability 1. 

{\it Remark:} It is worth mentioning here that (\ref{ARQR}) yields the maximum likelihood estimator (MLE) when the innovations $\delta_i$ from the model (\ref{QR}) - (\ref{AR}) are distributed as in (\ref{chexponential}).  Particularly in the case when $\tau = 0.5$, (\ref{chexponential}) becomes the Laplace distribution.  This analogy provides statistical safeguards to the solution in (\ref{ARQR}). In particular, although the density is not differentiable with respect to the parameter for all $x$, \cite{Daniels1961} proves a general theorem not requiring differentiability that establishes large sample properties (especially asymptotic normality) of the maximum likelihood estimator. Note also that \cite{Cramer1946a, Cramer1946} establishes consistency of such estimators under very general conditions (see also \cite{Lehmann1998} for more details).

As a result of Lemma \ref{lemma}, standard tools from convex optimization are not applicable for solving the above problem.  We therefore propose an iterative method to optimize the objective function.  The procedure we propose for obtaining parameter estimates for the model (\ref{QR}) - (\ref{AR}) alternates between fitting QR (assuming \emph{i.i.d.} residuals) and fitting an AR model to the residuals.  We shall refer to this as the ``QUAntile Regression with Time Series errors" algorithm (QUARTS). The method does assume however that the order of the residual AR process is given a priori.  This problem will be tackled after presenting the algorithm.  We now proceed to describe the QUARTS approach in detail.

Let $QRfit(\boldsymbol{y},\boldsymbol{X},\tau)$ denote the function that takes as arguments a vector $\boldsymbol{y}$ of $n$ responses, an $n \times (p+1)$ design matrix $\boldsymbol{X}$, and a quantile $\tau \in (0,1)$, and returns a column vector consisting of a fitted intercept followed by $p$ fitted regression coefficients obtained by applying some QR fitting technique to regress the conditional $\tau^{th}$ quantile of $\boldsymbol{y}$ on $\boldsymbol{X}$.  Furthermore, for an AR($q$) model of a time series $\{\varepsilon_i\}$

\begin{equation}
\label{ARint}
\varepsilon_i = \alpha + \phi_1 \varepsilon_{i-1} + \cdots + \phi_q \varepsilon_{i-q} + \delta_i,
\end{equation}
where $\alpha$ is an intercept term and the $\delta_i$ are \emph{i.i.d.} innovations, define the AR polynomial as the $q^{th}$-order polynomial $\phi(z) = 1- \phi_1 z - \cdots - \phi_q z^q$, and define the row vector $\boldsymbol{\phi} = (\phi_1, \dots, \phi_q)$.  Then the fitted model above for the residuals can be written as follows:
\begin{equation*}
\phi(B) \varepsilon_i = \alpha + \delta_i,
\end{equation*}
where $B$ denotes the backshift operator $B \varepsilon_i = \varepsilon_{i-1}$.  Now let $\boldsymbol{X}_i$ denote the $i^{th}$ row of the matrix $\boldsymbol{X}$, and for a row vector $\boldsymbol{x}^\top$, define the row vector $\phi(B) \boldsymbol{x}^\top$ coordinatewise, which is to say, $(\phi(B) \boldsymbol{x}^\top)_j = \phi(B) x_j$ .  Let $\tilde{\boldsymbol{X}}$ be the matrix composed of the lower $n-q$ rows of $\boldsymbol{X}$ (to be explained later).  With the above notation established, the QUARTS algorithm for fitting the conditional $\tau^{th}$ quantile for a given data set $\boldsymbol{y}$, $\boldsymbol{X}$ with AR($q$) residuals is given in Algorithm \ref{alg1}.

\begin{algorithm}
\caption{Quantile Regression with Time Series Errors (QUARTS)}
\label{alg1}
\begin{algorithmic}[1]
\REQUIRE $\boldsymbol{y}$, $\boldsymbol{X}$, $q$, $\tau$
\STATE Set $\boldsymbol{\phi}^{(0)} = \boldsymbol{0}^\top, \quad \boldsymbol{\varepsilon}^{(0)} = \boldsymbol{0}$.
\STATE  Given $\boldsymbol{\phi}^{(j-1)}, \boldsymbol{\varepsilon}^{(j-1)}$, let 
\[ \check{y}_{i-q} = y_i - \phi^{(j-1)}_1 \varepsilon^{(j-1)}_{i-1} - \cdots - \phi^{(j-1)}_q \varepsilon^{(j-1)}_{i-q}, \quad \forall i \in \{q+1,\dots,n\}$$$$
\boldsymbol{\beta}_{\tau}^{(j)} = QRfit(\check{\boldsymbol{y}},\tilde{\boldsymbol{X}},\tau)$$$$
\boldsymbol{\varepsilon}^{(j)} = \boldsymbol{y} - \boldsymbol{X} \boldsymbol{\beta}_{\tau}^{(j)} \]
\STATE Using QR, fit the regression model $\varepsilon^{(j)}_i = \phi_1 \varepsilon^{(j)}_{i-1} + \cdots + \phi_q \varepsilon^{(j)}_{i-q} + \delta_i$, for $i \in \{q+1,\dots,n\}$.  
Let $\boldsymbol{\phi}^{(j)}$ be the estimated AR row vector. 
\STATE Repeat steps 2 and 3 until $\boldsymbol{\phi}^{(j)}$ and $\boldsymbol{\beta}_{\tau}^{(j)}$ converge to steady-state solutions $\boldsymbol{\phi}$ and $\boldsymbol{\beta}_{\tau}$, respectively.  Return $\boldsymbol{\beta}_{\tau}$ as the final coefficient vector and $\boldsymbol{\phi}$ as the final AR model coefficient vector.
\end{algorithmic}
\end{algorithm}

The QUARTS algorithm can be framed as a cyclical block coordinate descent method where the two blocks correspond to regression parameters and autoregressive coefficients respectively. Such optimization methods applied to non-differentiable, non-convex functions have in general been shown to have good convergence properties. In particular, \citep{Tseng2001} proves that the cluster point of the sequence of iterates  given by the block coordinate descent method is a stationary point of the objective function (see also \cite{Sargan1964} for more details). The idea of the QUARTS algorithm is to iterate back and forth between fitting the QR regression coefficients and fitting the AR coefficients.  Once QR regression coefficients are obtained in step 2, AR coefficients are fit to the residuals in step 3.  When we iterate back to step 2, the deterministic AR component of the residuals, i.e., $\phi_1 \varepsilon^{(j)}_{i-1} + \cdots + \phi_q \varepsilon^{(j)}_{i-q}$ for the $i^{th}$ residual, is then removed from the response vector $\boldsymbol{y}$ to generate a new vector $\check{\boldsymbol{y}}$. New QR coefficients are then obtained using $\check{\boldsymbol{y}}$ as the response, and we return again to step 3.  Note that we choose to fit only the latter $n-q$ data points so as to avoid edge effects from the first $q$ points, for which the full AR model (\ref{ARint}) is not defined because it requires the previous $q$ residuals.  

\subsection{Choosing a Residual Autoregressive Order}
As described in the previous subsection, given some QR fitting procedure \emph{QRfit}, the order of the AR residual process $q$ is required a priori before using the QUARTS algorithm.  The value of $q$ can be determined through a combination of calculating residuals and statistical hypothesis testing to assess serial correlation. This process of determining the lag of the AR process is described in Algorithm \ref{alg2}.

\begin{algorithm}
\caption{Residual Autoregressive Lag Determination Algorithm (RARLD)}
\label{alg2}
\begin{algorithmic}[1]
\STATE Set $q=0$
\STATE Given $q$, fit the data with QUARTS and let the innovations be $\delta_1, \dots, \delta_n$.
\STATE If the time series $\delta_1, \dots, \delta_n$ exhibits AR behavior, set $q = q+1$ and return to step 2.  Otherwise, return $q$.
\end{algorithmic}
\end{algorithm}

Note that in the first pass of this algorithm, QUARTS need not be applied since $q=0$ corresponds to assuming \emph{i.i.d.} residuals.  In other words, standard QR fitting is sufficient. The RARLD algorithm above starts with a lag of $q=0$ and successively increases $q$ until the innovations no longer exhibit AR behavior.  In particular, the statistical hypothesis testing component comes in at step 3, in deciding whether or not the innovations exhibit AR behavior.  For these purposes, one can employ the autocorrelation function (ACF), partial autocorrelation function (PACF), and Ljung-Box tests up to some given lag.  Using test levels of 5\% to evaluate the significance of these tests, statistically informed assessments of the AR structure (or lack thereof) could be made for a given set of innovations.

\subsection{Statistical Inference for Autoregressive Quantile Regression Outputs}
The previous subsections (2.2 and 2.3) define an objective function and propose an iterative method to determine parameter estimates.  For this purpose, the distribution of the innovations does not need to be specified.  We now proceed to develop a statistical framework for uncertainty quantification of parameter estimates, which on the other hand does require the specification of a statistical model for the innovations. This is akin to least squares minimization in OLS vs. specifying the distribution of the $\varepsilon_i \iid N(0, \sigma^2)$ for the purposes of statistical inference.  In particular, there are three main model outputs for which we are interested in quantifying uncertainty: the regression coefficients $\hat{\boldsymbol{\beta}}_{\tau}$ and $\hat{\boldsymbol{\phi}}_{\tau}$, the conditional quantiles $\hat{Q}^{(\tau)}(y_i)$, and the reconstructed temperatures (or hindcasts) $\hat{y}_i$ (which we shall see will have the same estimates as the $\hat{Q}^{(\tau)}(y_i)$, but with different uncertainty).  For all three we assume that a good approximation to the distribution of the residual innovations can be made, so that a form of parametric bootstrap can be employed (although our methodology is flexible enough to allow for the non-parametric bootstrap).  

{\bf \emph{A. Estimating the innovation distribution}}: Suppose we are given a data set consisting of a response vector $\boldsymbol{y} = (y_1,\dots,y_n)^\top$ and predictor variables $\boldsymbol{x}_1,\dots,\boldsymbol{x}_n \in \mathbb{R}^p$.  Recall from Section 2.2:
\[ y_i = \boldsymbol{x}_i^\top \boldsymbol{\beta}^{(\tau)} + \varepsilon_i^{(\tau)}, $$$$
\varepsilon_i^{(\tau)} = \phi_1^{(\tau)} \varepsilon_{i-1}^{(\tau)} + \cdots + \phi_q^{(\tau)} \varepsilon_{i-q}^{(\tau)} + \delta_i^{(\tau)}, \quad \delta_i^{(\tau)} \iid f_{\delta}^{(\tau)}, \quad s.t. \quad \mathbb{P}_{f_{\delta}^{(\tau)}}(\delta_i < 0) = \tau, $$$$
Q^{(\tau)}(y_i) | \boldsymbol{x}_i = \boldsymbol{x}_i^\top \boldsymbol{\beta}^{(\tau)} + \phi_1^{(\tau)} \varepsilon_{i-1}^{(\tau)} + \cdots + \phi_q^{(\tau)} \varepsilon_{i-q}^{(\tau)}. \]

After using a fitting method such as QUARTS to generate parameter estimates $\hat{\boldsymbol{\beta}}^{(\tau)}$ and $\hat{\phi}_1^{(\tau)},\dots,\hat{\phi}_q^{(\tau)}$, the conditional $\tau^{th}$ quantile of the $n$ observations can be estimated by setting $\hat{\varepsilon}_{-q+1}^{(\tau)},\dots,\hat{\varepsilon}_0^{(\tau)} = 0$ and letting
\begin{equation}
\label{condQ}
\hat{Q}^{(\tau)}(y_i) | \boldsymbol{x}_i = \boldsymbol{x}_i^\top \hat{\boldsymbol{\beta}}^{(\tau)} + \hat{\phi}_1^{(\tau)} \hat{\varepsilon}_{i-1}^{(\tau)} + \cdots + \hat{\phi}_q^{(\tau)} \hat{\varepsilon}_{i-q}^{(\tau)},
\end{equation}
where $\hat{\varepsilon}_i^{(\tau)} = y_i - \boldsymbol{x}_i^\top \hat{\boldsymbol{\beta}}^{(\tau)}$ for $i \in \{1,\dots,n\}$.  Note that the first $q$ estimates will be directly tainted by edge effects from setting $\hat{\varepsilon}_{-q+1}^{(\tau)},\dots,\hat{\varepsilon}_0^{(\tau)} = 0$, and can be discarded, or if retained, should be used with little confidence.  Discarding the first $q$ observations, we now have a sample of size $n-q$ of estimated innovations, calculated as
\[ \hat{\delta}_i^{(\tau)} = y_i - \boldsymbol{x}_i^\top \hat{\boldsymbol{\beta}}^{(\tau)} - \hat{\phi}_1^{(\tau)} \hat{\varepsilon}_{i-1}^{(\tau)} - \cdots - \hat{\phi}_q^{(\tau)} \hat{\varepsilon}_{i-q}^{(\tau)}, \quad i \in \{q+1, \dots, n\} \]
from which to try and estimate $f_{\delta}^{(\tau)}$ for the purposes of uncertainty quantification. This can be achieved in a parametric way by assuming a distribution for $f_{\delta}^{(\tau)}$ and using the parametric bootstrap. Alternatively, the empirical distribution of the $\hat{\delta}_i^{(\tau)}$ can also be used to estimate $f_{\delta}^{(\tau)}$, which is equivalent to using the nonparametric bootstrap.

{\bf \emph{B. Uncertainty quantification of regression and AR coefficients}}: For regression coefficient uncertainty, we follow the standard parametric bootstrap procedure, except we hold the AR order of the residuals, $q$, as a constant across bootstrap replications.  Using $f_{\delta}^{(\tau)}$ to generate new \emph{i.i.d.} innovations $\tilde{\delta}^{(\tau)}$, we can create a new stationary time series for each bootstrap replication:
\begin{equation}
\label{bootAR}
\tilde{\varepsilon}_i^{(\tau)} = \hat{\phi}_1^{(\tau)} \tilde{\varepsilon}_{i-1}^{(\tau)} + \cdots + \hat{\phi}_q^{(\tau)} \tilde{\varepsilon}_{i-q}^{(\tau)} + \tilde{\delta}_i^{(\tau)}, \quad i \in \{1,\dots,n\},
\end{equation}
so that a new bootstrap response vector can be calculated as
\begin{equation}
\label{bootY}
\tilde{y}_i = \boldsymbol{x}_i^\top \hat{\boldsymbol{\beta}}^{(\tau)} + \tilde{\varepsilon}_i^{(\tau)}, i \in \{1,\dots,n\}.
\end{equation}
Then the bootstrapped coefficient estimates, denoted by the $\tilde{\boldsymbol{\beta}}^{(\tau)}$'s and $\tilde{\boldsymbol{\phi}}^{(\tau)}$'s, are each calculated by using the same method (QUARTS) as was used to calculate $\hat{\boldsymbol{\beta}}^{(\tau)}$ and $\hat{\boldsymbol{\phi}}^{(\tau)}$.  The distribution of these bootstrap coefficients can now be used to quantify the uncertainty of the QUARTS estimates for the parameters $\boldsymbol{\beta}$ and $\boldsymbol{\phi}$.

{\bf \emph{C. Out of sample conditional quantile estimation}}: On the other hand, for uncertainty of out-of-sample predictions, more care is needed.  First we need to discuss how to generate out-of-sample point estimates.  In particular, suppose that we wish to estimate the conditional $\tau^{th}$ quantile of the next $m$ realizations of the response variable, $y_{n+1},\dots,y_{m+n}$, based on the next $m$ realizations of the predictor variables, $\boldsymbol{x}_{n+1},\dots,\boldsymbol{x}_{n+m}$.  Due to the model specifications in (2.4) and (2.5), for any given $\tau$, the estimated quantiles $\hat{Q}^{(\tau)}(y_{n+1}) | \boldsymbol{x}_{n+1},\dots, \hat{Q}^{(\tau)}(y_{n+m}) | \boldsymbol{x}_{n+m}$ can also serve as point predictions of $y_{n+1},\dots,y_{m+n}$. However, to minimize uncertainty and bias, this is best done with $\tau = 0.5$: the conditional median.  This is analogous to using the estimated conditional mean, $\widehat{E[\boldsymbol{y}|\boldsymbol{x}]} = \hat{\boldsymbol{\beta}}_{LS} \boldsymbol{x}$, for out-of-sample prediction in the least squares setting (for the sake of brevity new notation is not introduced here).

For out-of-sample prediction, namely $i > n$, we no longer have $y_i$ with which to calculate $\hat{\varepsilon}_i^{(\tau)}$.  Using the (assumed known) distribution $f_{\delta}^{(\tau)}$ of the $\hat{\delta}_i^{(\tau)}$, the $\hat{\varepsilon}_i^{(\tau)}$ can be recursively built up as
\begin{equation}
\hat{\varepsilon}_i^{(\tau)} = \hat{\phi}_1^{(\tau)} \hat{\varepsilon}_{i-1}^{(\tau)} + \cdots + \hat{\phi}_q^{(\tau)} \hat{\varepsilon}_{i-q}^{(\tau)} + \mu_{\delta}^{(\tau)}
\label{ARpred}
\end{equation}
for $(i>n)$, where $\mu_{\delta}^{(\tau)}$ is some summary parameter of the distribution $f_{\delta}^{(\tau)}$ that represents the best estimate of the out-of-sample $\delta_i^{(\tau)}$ (for example, $\mu_{\delta}^{(\tau)}$ might be the mean of $f_{\delta}^{(\tau)}$ if $f_{\delta}^{(\tau)}$ is Gaussian).  Then the point predictions $\hat{y}_{n+1},\dots,\hat{y}_{m+n}=\hat{Q}_{n+1}^{(\tau)}(y_{n+1}) | \boldsymbol{x}_{n+1},\dots,\hat{Q}_{n+m}^{(\tau)}(y_{n+m}) | \boldsymbol{x}_{n+m}$ can be calculated from (\ref{condQ}).  

{\bf \emph{D. Uncertainty quantification of out-of-sample point predictions and conditional quantile estimates}}: Now that we have point predictions, we can estimate their parametric bootstrap sampling distributions for the purposes of uncertainty quantification.  As mentioned earlier, we assume that $f_{\delta}^{(\tau)}$ can be estimated from the sequence $\hat{\delta}_{q+1}^{(\tau)},\dots,\hat{\delta}_n^{(\tau)}$.  To estimate the pathwise distribution of the reconstruction $\hat{y}_{n+1},\dots,\hat{y}_{m+n}$, we will generate many such sample paths $\tilde{y}_{n+1},\dots,\tilde{y}_{m+n}$ using the parametric bootstrap and look at the resulting empirical distribution of such paths.  Each path is generated as follows:

Use $f_{\delta}^{(\tau)}$, $\hat{\boldsymbol{\beta}}^{(\tau)}$, and $\hat{\boldsymbol{\phi}}^{(\tau)}$ to generate a bootstrap response vector $\tilde{y}_1,\dots,\tilde{y}_n$ as in (\ref{bootAR}) and (\ref{bootY}).  From $\tilde{y}_1,\dots,\tilde{y}_n$ and $\boldsymbol{x}_1,\dots,\boldsymbol{x}_n$, generate bootstrap coefficient estimates $\tilde{\boldsymbol{\beta}}^{(\tau)}$ and $\tilde{\boldsymbol{\phi}}^{(\tau)}$, and calculate the bootstrap path recursively as in (\ref{ARpred}):
\[ \tilde{\varepsilon}_i^{(\tau)} = \tilde{\phi}_1^{(\tau)} \tilde{\varepsilon}_{i-1}^{(\tau)} + \cdots + \tilde{\phi}_q^{(\tau)} \tilde{\varepsilon}_{i-q}^{(\tau)} + \tilde{\delta}_i^{(\tau)}, \quad i \in \{n+1,\dots,n+m\} \]
\[ \tilde{y}_i = \boldsymbol{x}_i^\top \tilde{\boldsymbol{\beta}}^{(\tau)} + \tilde{\varepsilon}_{i}^{(\tau)}, \quad i \in \{n+1,\dots,n+m\} \]
with the $\tilde{\delta}_i^{(\tau)}$ drawn \emph{i.i.d.} from $f_{\delta}^{(\tau)}$.  Similarly, for measuring uncertainty in the conditional quantile estimates, we can calculate bootstrap paths of the conditional quantiles as
\[ \tilde{Q}_i^{(\tau)}(y_i) | \boldsymbol{x}_i = \boldsymbol{x}_i^\top \tilde{\boldsymbol{\beta}}^{(\tau)} + \tilde{\phi}_1^{(\tau)} \tilde{\varepsilon}_{i-1}^{(\tau)} + \cdots + \tilde{\phi}_q^{(\tau)} \tilde{\varepsilon}_{i-q}^{(\tau)} = \tilde{y}_i - \tilde{\delta}_i^{(\tau)}, \quad i \in \{n+1,\dots,n+m\} \]
Note that the difference between out-of-sample point prediction, uncertainty quantification of this predicted value, and uncertainty quantification of the conditional quantile estimate lies in either the use of $\mu_{\delta}^{(\tau)}$, the use of $\tilde{\delta}_i^{(\tau)}$, or the exclusion of the $\tilde{\delta}_i^{(\tau)}$, respectively.  This distinction is analogous to the difference between predictions, prediction intervals, and confidence intervals in the traditional Gaussian least squares setting. By using the above process to quantify uncertainty, we simultaneously take into account both coefficient uncertainty (including coefficient interdependencies) and the uncertainty from the random $\delta^{(\tau)}$'s.  We note that the above approach extends the uncertainty quantification technique used in \cite{Li2007}.

{\bf \emph{E. Overfitting of innovation distribution}}: A quick note on estimating $f_{\delta}$ is in order.  Any parameters associated with the standard deviation of $f_{\delta}$ should be approached with caution.  For example, choosing $f_{\delta} \sim N(\hat{\mu},\hat{\sigma}^2)$, where $\hat{\mu}$ is the sample mean and $\hat{\sigma}$ the sample standard deviation of $\hat{\delta}_{q+1}^{(\tau)},\dots,\hat{\delta}_n^{(\tau)}$, would constitute overfitting.  This reasoning follows from the fact that the model is fit by solving an optimization problem that minimizes an analogue of the variability parameter that is similar to $\hat{\sigma}$ (namely the average check function of the innovations).  An improvement would be to use a method similar to cross validation that is used in \cite{Li2007}: which is to split up the data into sections, and for each section fit the model to all the other sections and calculate the sample standard deviation of the prediction innovations from that model within that section. Thereafter the average of these sample standard deviations across sections is used as a value for $\hat{\sigma}$.  It is clear that this holdout procedure aims to reduce overfitting.

\section{Background on Paleoclimate Data and Implementation}
\subsection{Overview of Paleoclimate Proxy Data}
\begin{figure}[h!]\centering
\includegraphics[scale=.5]{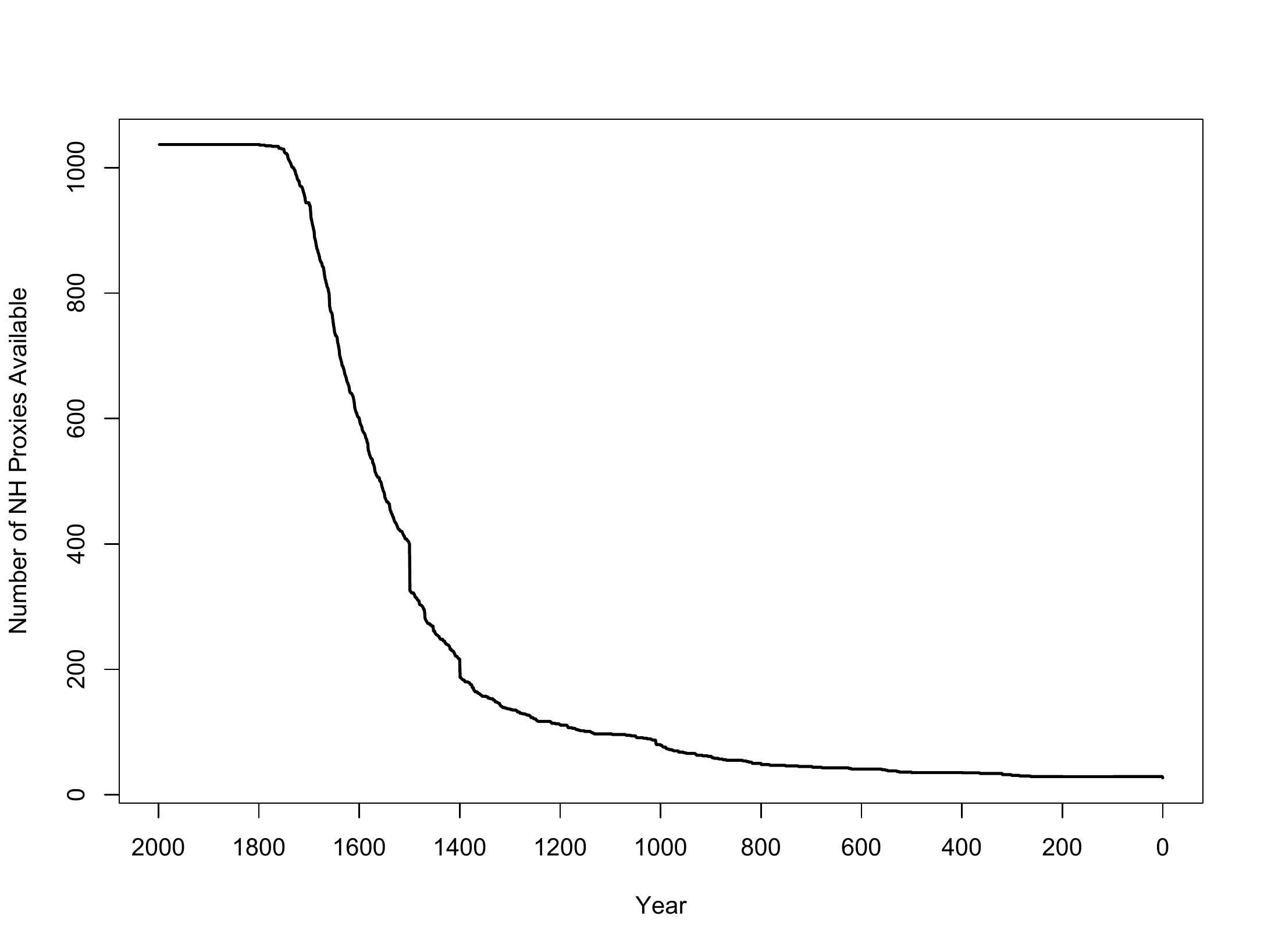}
\caption{Number of M08 Northern Hemisphere proxy records available by year from 1998 AD to 0.}
\label{ProxyAvailTime}
\end{figure}

\begin{figure}\centering
\includegraphics[scale=.41]{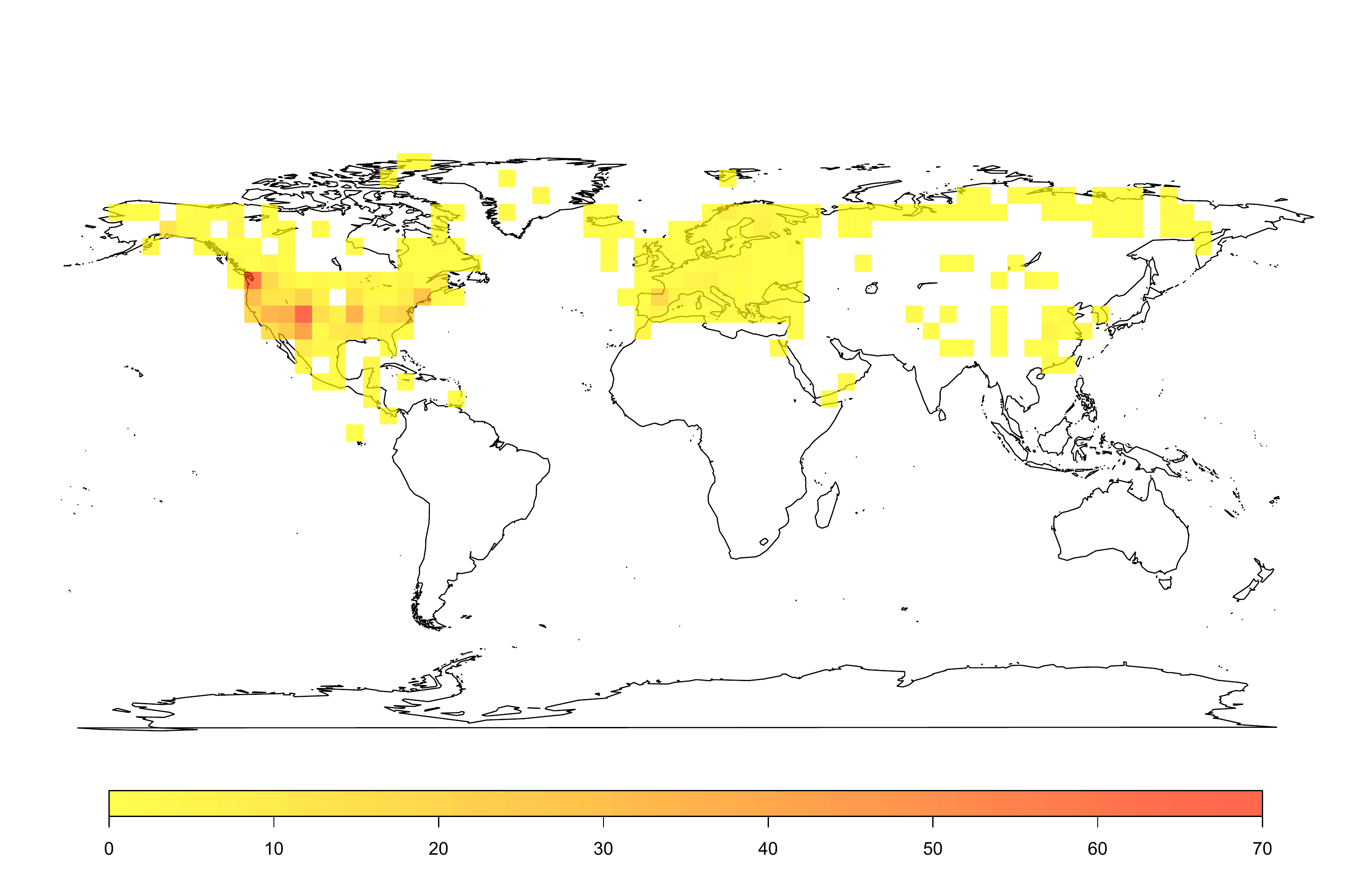}
\caption{Number of M08 Northern Hemisphere proxy records available by location on a 5\textdegree{}$\times$5\textdegree{} grid.  Only grid-points with at least one proxy are colored.}
\label{ProxAvailLoc}
\end{figure}

A widely used paleoclimate proxy data set is that found in \cite{Mann2008} (heretofore M08\footnote{Available at: \texttt{http://www.meteo.psu.edu/\textasciitilde mann/supplements/MultiproxyMeans07/data/}}), and will also be used for illustrating the methodology in this paper.  The proxy record in M08 contains 1,209 time series, 1,037 of which are in the northern hemisphere (NH), and is comprised of tree-ring, marine sediment, speleothem, lacustrine, ice core, coral, and historical documentary series \citep{Mann2008}.  The NH proxy series span 248 5\textdegree{}$\times$5\textdegree{} grid-points on the globe.  They are given at annual resolution, with the vast majority beginning between the years 1000 and 1800, and all of them ending between 1998 and 2003.  Figures \ref{ProxyAvailTime} and \ref{ProxAvailLoc} show the number of available proxies over time and space, respectively.

M08 also contains a number of instrumental records, originating from the University of East Anglia (Norwich, UK) Climatic Research Unit\footnote{Raw data used is `HadCRUT3v' from: \texttt{http://www.cru.uea.ac.uk/cru/data/temperature/}}.  The instrumental record we use from M08 is a spatial average of annual NH land and ocean temperatures, for which missing spatial points are imputed using the RegEM method \citep{Schneider2001}.  This series runs from 1850 to 2006 and has had the mean of the temperature during the base period 1961-1990 subtracted off.  Both the instrumental and proxy records have had substantial imputation and smoothing.  Like many others before, for the purposes of this paper we take this data as given and assume that the aforementioned imputations/smoothing were performed properly.

\subsection{Implementation of QUARTS Methodology in the Proxy Data Context}
Section 2 was intentionally made as general as possible in order to develop the statistical methodology in this paper.  In the current subsection, we will go into more specifics of the paleoclimate reconstruction application, recalling that the methodology was motivated by this particular application.  In particular, by using a robust modelling procedure, we aim to assess if we can reduce the reconstruction uncertainty reported in previous work \citep{Li2007, Mann2008, McShane2011}. Having said this, we observe that the statistical methodology that is developed in Section 2 can also be useful in other application areas.

It is important to also point out the assumptions that are implicit in our model.  Regarding the data itself, as mentioned above, the available instrumental and proxy record is taken for granted, although it is far from being a collection of direct observations of response and predictor variables.  For the model, we assume that the relationship between the proxies and the conditional temperature quantiles is linear (or at least can be well approximated by a linear relationship over the range of values that appear in the time-span of our reconstruction).  Note that this assumption is a relaxation of the explicit assumption of normal errors that is made in any standard LS paleoclimate reconstruction.  A further assumption that is implicitly imposed is the principle of uniformitarianism, which is that the relationship between the proxies and temperature is constant over time.

There are a few other implementation details that are specific to the paleoclimate proxy application under consideration.  For the sake of completeness, we will elaborate on the challenges that arise in this regard in the remainder of this section, and present the results of our two analyses in the following two sections.  A reader interested in the final results can skip this subsection and proceed straight to Section 4.

{\bf \emph{High dimensional nature of paleoclimate data sets}}: We first note that the number of instrumental data points is small compared to the number of paleoclimate proxy series, and thus, the paleoclimate problem is very much in the domain of the \emph{large $p$, small $n$} regime.  Paleoclimate reconstruction problems often require some form of regularization, and we will use principal components quantile regression (PCQR) for this purpose.  PCQR mimics its LS counterpart by first rotating a problem's design matrix from its original predictor-space to its eigenspace. Thereafter, only the first $k$ components are retained, ordered by decreasing eigenvalue, before the fitting technique is applied. In particular, principal components is used to reduce the number of proxies, and the principal components themselves now serve as the predictor variables.  Ten-fold cross validation \citep{Hastie2009} is used to choose the number of principal components, $k$, that are retained.  The choice of using principal components, as opposed to other regularization methods that penalize the magnitude of the regression coefficients (in particular L1-penalization/Lasso and L2-penalization/Ridge), was preferred for our approach because penalties result in biased estimates. The standard bootstrap has more difficulty accounting for the extra error that comes from biased estimation.  For the remainder of the paper, QUARTS will refer to the same algorithm when dimension reduction via principle components is incorporated.

{\bf \emph{Non-i.i.d. residuals}}: Autocorrelation in the residuals prevents random allocation of data points to folds for the purposes of cross validation. In particular, standard cross validation is often implemented by splitting a data set randomly into folds and for each fold, fitting the model to the rest of the data and measuring the predictive power of that model on the holdout fold.  However, random allocation to folds from our dataset would destroy the time series structure of the data, and thus the data is instead split evenly into ten contiguous blocks.  For the first and last holdout blocks, the model is fit to the remaining nine blocks as usual, but for the middle eight blocks, the edge effects mentioned in Section 2.4 need to be accounted for twice in order to account for the break between the two blocks of data (to be discussed later in this subsection).  

Another challenge arises because the goal of QUARTS is to minimize the value of 
\begin{equation}
\label{checkError}
\rho_{\tau} (y_i - \boldsymbol{x}_i^{\top} \hat{\boldsymbol{\beta}} - \hat{\phi}_1 \hat{\varepsilon}_{i-1} - \cdots - \hat{\phi}_q \hat{\varepsilon}_{i-q})
\end{equation}
over all observations $(y_i,\boldsymbol{x}_i^{\top})$.  However, although the $\boldsymbol{x}_i^{\top}$ are always available, $y_i$ is naturally not available out-of-sample (i.e., in a given holdout block), and thus any out-of-sample $\hat{\varepsilon}_i$ cannot be computed exactly as $y_i - \boldsymbol{x}_i^{\top} \hat{\boldsymbol{\beta}}$, but need to be estimated from their time series structure using (\ref{ARpred}).  However, estimated in this manner, the out-of-sample $\hat{\varepsilon}_i$ will quickly converge to a mean value, losing all time series structure.  Thus performing a truly out-of-sample prediction on the holdout block does not adequately incorporate the AR component of the fitted model.  To solve this problem, the model is instead propagated backwards through the validation period using the true residuals as calculated using the known temperatures.  Thus we are explicitly using the check function of the innovations as suggested by (\ref{checkError}) to measure error.

{\bf \emph{Strong autoregressive structure in temperature record}}: We note that the temperature record itself has an AR structure. This property encourages a stronger AR component in the model, though the goal is for the proxy variables to explain the response variable as much as possible.  A time series model with no predictors fits the temperature record fairly well, but has very little predictive power in a hindcast.  This becomes a problem when assessing the fit of models with a very low number of principal components ($k$), because cross validation will tend to favor them as having small out-of-sample error.  In order to mitigate this effect, the models tested by cross validation were restricted to have at least $k=3$ components.

{\bf \emph{Edge effects}}: The cross validation holdout block at the beginning of the sample will have its first few points tainted by the edge effects from the AR structure of the residuals.  In using (\ref{checkError}) to measure error in the holdout blocks for validation, it is required that the preceding $q$ residuals are defined for each point.  This is not the case for the first holdout block (to be more precise, the holdout block at the end of the sample in our case, since we fit backwards in time), and this edge effect taints the first few error measurements.  To correct for this, instead of splitting the entire instrumental period of 1850-1998 into ten blocks, only 1850-1994 was split.  For the holdout block ending in 1994, the period 1995-1998 is excluded from the calibration and validations periods, but instead is used to absorb the aforementioned edge effect.  For the other cross validation holdout blocks, the years 1995-1998 is included in the calibration period as usual.  The choice of four points to absorb the edge effect represents a compromise between removing too many points and keeping the validation measurements in the last block uncorrupted, and is also justified because $q$ is very small.

{\bf \emph{Innovation distribution overfitting}}: A problem that is often encountered is that using the in-sample variance of the estimated innovations leads to overfitting.  We will see in the following two sections that Gaussian approximations $N(\mu,\sigma^2)$ for $\hat{f}_{\delta}$ will be adequate for uncertainty quantification.  For $\hat{\mu}$ we can use the sample mean, but to avoid overfitting, we adapted a technique from \cite{Li2007} for estimating the variance.  We split the instrumental period minus four years into ten contiguous parts, and for each part, fit the model to the rest of the data. The true residuals were then used to propagate that model through the holdout block (so far the same process as the one used by cross validation to choose the number of principal components), then calculated the sample standard deviation (dividing by $n-1$ instead of $n$ to remove bias) of the innovations.  These ten values of standard deviation were then averaged to obtain our estimate $\hat{\sigma}$ of the standard deviation of the true innovation distribution.

For the sake of completeness, one thing should be noted about this last technique.  Although the goal is to avoid overfitting, the method proposed above does not quite reach that goal.  By taking the number of principal components and the AR order of the residuals as given, model features are included that have already been fit to the whole data set, and thus the ten models fit are not \emph{truly} out-of-sample.  However, the alternative is to use cross validation to choose the number of principal components for each of the ten holdout blocks (which introduces unacceptably high variance into the model) and to go through the RARLD algorithm for each cross validation sample, which in turn is prohibitively slow.  Furthermore, there is reason to believe the overfitting should be fairly slight: only the number of principal components, and not the principal components themselves, have been fit to the data.  In addition, components are included in the model in order of their eigenvalues, an ordering which is completely independent of the response vector.  For the same reasons above, the same assumption (holding the number of principal components and the residual AR order constant) is made in all bootstrapping calculations as well.  Moreover, since each bootstrap data set is generated using the same AR order as the original model, it is likely that this same order would be chosen anyway. Hence allowing for uncertainty in the lag will tend to have less merit when carrying out bootstrap calculations. 

\section{Robust Millennial Reconstruction with Reduced Uncertainty}
We now proceed to apply the methodology developed in previous sections to an important problem in the field of climate change, namely the task of reconstructing past climates. As compared to least squares methods, least absolute deviation methods are quite robust, suggesting the use of QUARTS at the $0.5^{th}$ quantile to reconstruct a conditional median.  This is comparable to, but promises to be more robust than, previous reconstructions of the conditional mean.  Many of the proxies, as pointed out in \cite{McShane2011}, are highly irregular time series. Hence, the relationship between the proxies and temperature may be quite different from the restrictive setting in which least squares regression is asymptotically efficient.

For the millennial reconstruction, we choose 998 AD as the start year, which is one millennium before the earliest of the proxy end years, 1998.  Since not all the proxies go back to 998, the reconstruction uses only the 79 proxies that do. Although this represents a fairly restricted subset of the entire proxy record, the authors are not aware of any systematic effect such a selection might have on the reconstruction or its uncertainty.  We note that our approach in this paper can be easily extended to take advantage of all the available paleoclimate proxy records. In particular, the missing values in the proxy record can be imputed using the expectation-maximization (EM) algorithm \citep{Schneider2001}. This approach allows QUARTS to be implemented on all 1,037 NH proxies. The final conditional median model that was fitted uses 9 principal components with AR(1) residuals.  Appendix B gives the details of determining the lag of the AR residual process through the RARLD algorithm, fitting $\hat{f}_{\delta}$, and correcting for overfitting. $\mathsf{R}$ code to implement the QUARTS methodology is available at...

\begin{figure}[ht!]\centering
\includegraphics[scale=.45]{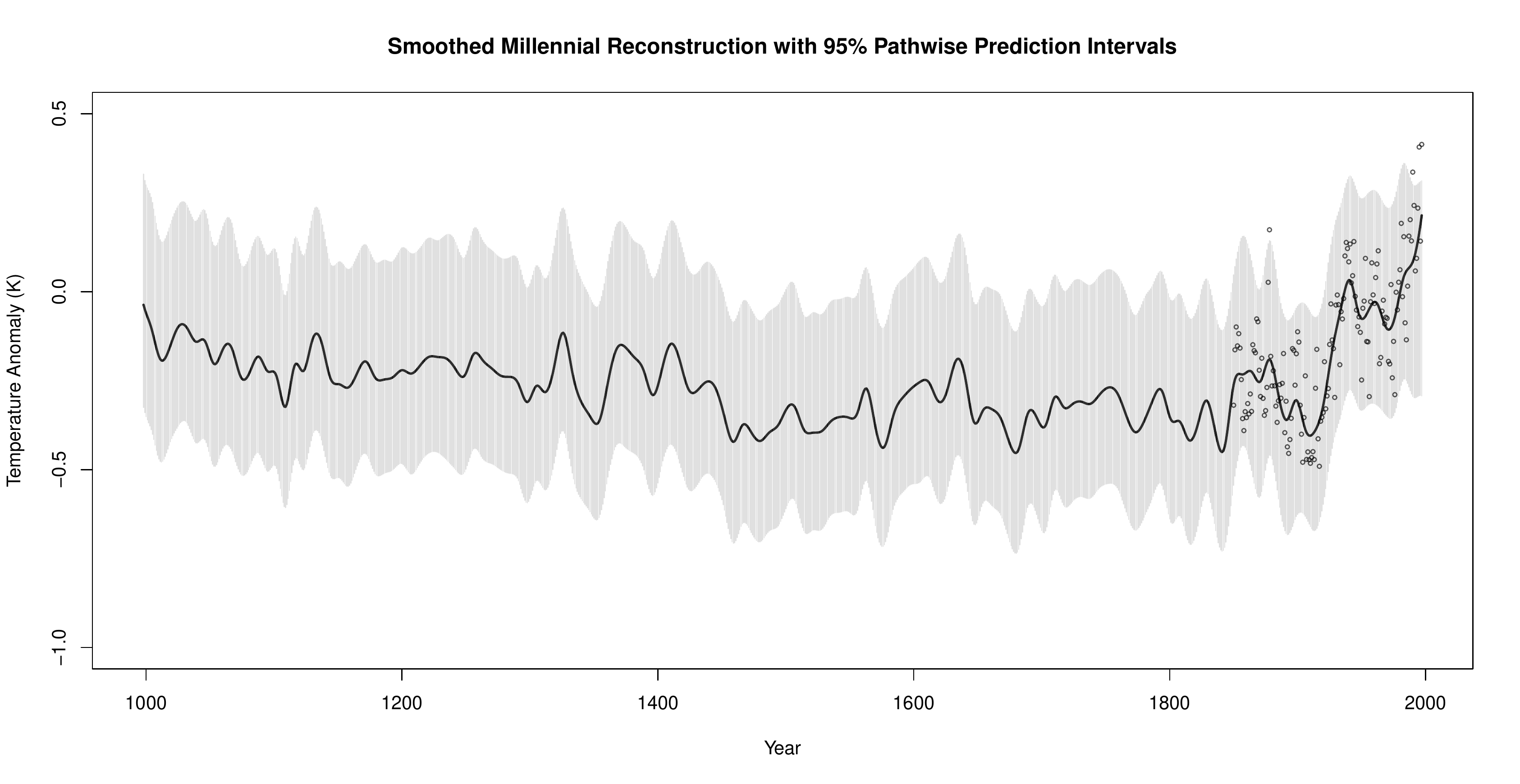}
\caption{Smoothed QUARTS temperature reconstruction (with 95\% pathwise prediction intervals) for the period 998 - 1997.  Instrumental temperatures are plotted as black circles for the period 1850 - 1997. Units are in Kelvin (K).}
\label{79_QRreconstruction}
\end{figure}

Figure \ref{79_QRreconstruction} shows the 1000-year reconstruction, along with a 95\% parametric bootstrap pathwise confidence interval. The reconstruction and its prediction interval endpoints are smoothed using a cubic spline with 115 degrees of freedom. Note that the uncertainty is calculated as in part D of subsection 2.4, i.e., it is prediction uncertainty, where the conditional median provides the point estimates.  This is entirely analogous and comparable to prediction uncertainty using the conditional mean for point prediction in the standard least squares setting.
The QUARTS reconstruction has several attractive features.  First, the reconstruction is qualitatively similar to previous reconstructions in the sense that the general ``hockey stick" shape is still apparent.  We also note that for the beginning of the reconstructed millenium, the QUARTS reconstruction is more similar to earlier reconstructions \citep{Mann1998, Mann2008} than that in \cite{McShane2011} in that it does not run up as high in the beginning of the reconstructed period. 

To see improvements in uncertainty, it can be compared directly to an analogous analysis (choosing the AR residual order, number of principal components, and correcting for overfitting in the same way) using generalized least squares (GLS).   Although the authors do not explicitly present a reconstruction, this is exactly the method used in \cite{Li2007} applied to the chosen number of principal components (M08 was not available in 2007, and thus only 14 proxies were used in their reconstruction, obviating the need for regularization).  For GLS, AR(0) and AR(1) residual structures are rejected while an AR(2) residual structure, with 3 principal components, does not reject the \emph{i.i.d.} innovations hypothesis, so this is the model chosen for the GLS fit.  Already, an advantage of the QUARTS approach are revealed: GLS fits best with fewer principal components and a higher order residual AR structure, representing a shift in the model away from the proxies and towards the residuals as compared to QUARTS.  Such a shift means that the GLS model will lose relatively more explanatory power going back in time, as the AR residual structure falls off quickly in magnitude as one leaves the instrumental period and reconstructs years prior to 1850.  This follows from the fact that the residuals cannot be directly computed out-of-sample where the response variable is unavailable and thus must be estimated from the model (\ref{AR}), and without the randomness of the $\delta_i^{(\tau)}$, (\ref{AR}) quickly converges to a constant.

\begin{figure}[ht!]\centering
\includegraphics[scale=.45]{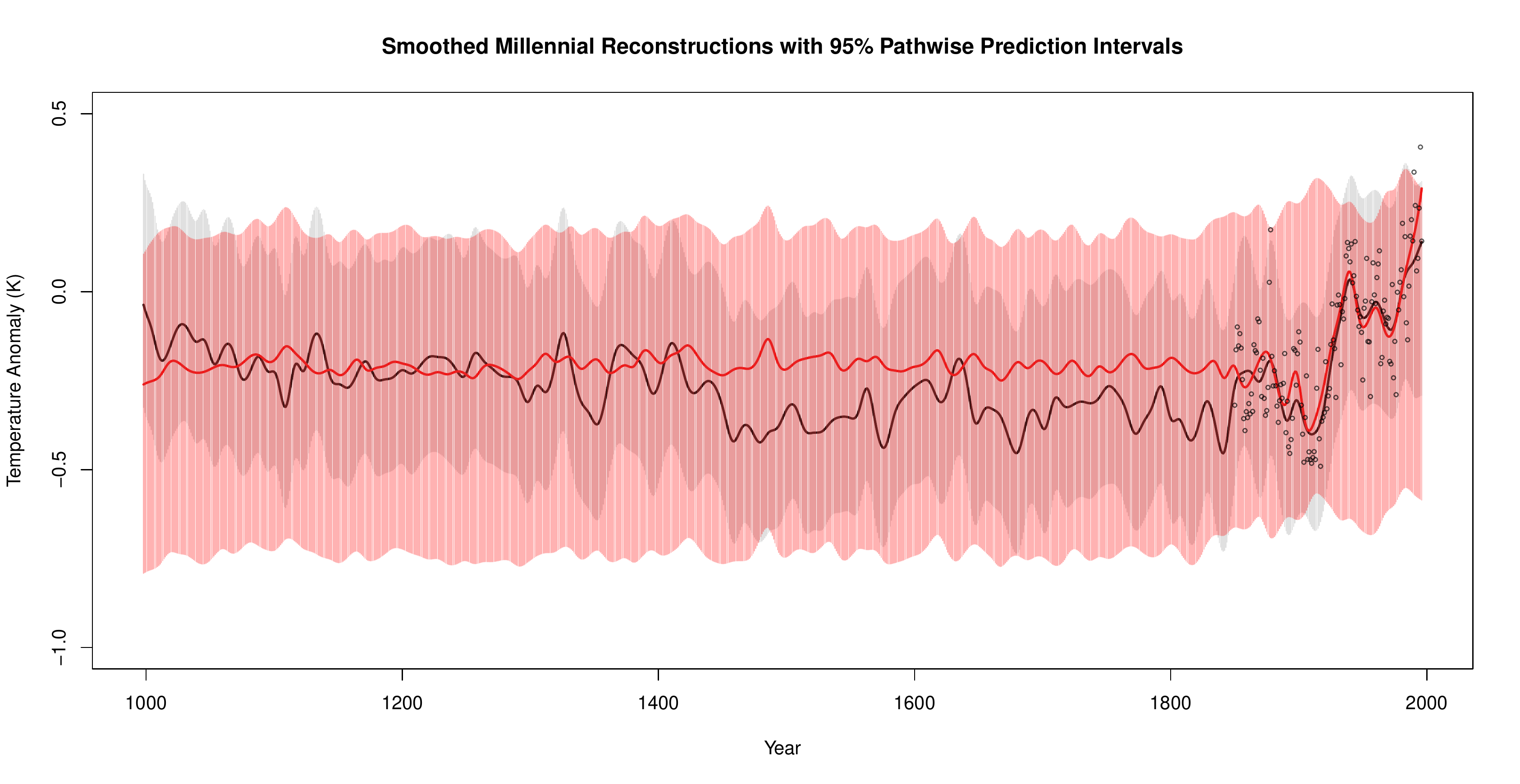}
\caption{Smoothed QUARTS (black) and GLS (red) temperature reconstructions (with 95\% pathwise prediction intervals) for the period 998 - 1996.  Instrumental temperatures are plotted as black circles for the period 1850 - 1996.}
\label{79_TWOreconstructions}
\end{figure}
Figure \ref{79_TWOreconstructions} compares the two reconstructions (both using the same smoothing technique as in Figure \ref{79_QRreconstruction}). The improvement in prediction interval width is quite stark.  The average out-of-sample prediction interval widths are 0.61 \textdegree C for QUARTS vs. 0.91 \textdegree C for GLS.  There is thus a 50\% increase in uncertainty in the GLS approach. The QR reconstruction also appears to capture the comparatively cool period just before the industrial revolution, known as the ``Little Ice Age" \citep{Matthes1939, Lamb1990, McShane2011} better than GLS.  For both models, the in-sample uncertainty is shown for reference. Since the temperature during the instrumental period is given and therefore fixed, there is no uncertainty in this period.  In this case, the in-sample uncertainty is just the fitted values with the innovation variance corrected for overfitting.  As expected, the actual coverage in-sample is slightly higher than 95\% (to account for overfitting) for both models.  It should also be noted that although applying the RARLD algorithm to GLS results in an AR(2) residual structure, as a heuristic lower bound on the GLS uncertainty, we also considered a GLS model with the more favorable values of $q=1$ and $k=9$. Note these values for $q$ and $k$ come from the QUARTS fit and are entirely artificial and optimistic for GLS, which chooses $q=2$ and $k=9$.  Fixing these parameters for the GLS fit results in \emph{i.i.d.} innovations and produces a point reconstruction similar to the two shown in Figure \ref{79_TWOreconstructions} with an average out-of-sample prediction interval width of 0.70 \textdegree C -- still non-negligibly larger than that of QUARTS.

We can also compare the two models by looking at coefficient p-values.  Using the principal component transformation, principal component coefficients can be converted to proxy coefficients, and bootstrap p-values for the proxy coefficients are easily calculated following part B of subsection 2.4. Table \ref{tab79pval} compares the numbers of significant proxy coefficients for the 79 proxies available going back to 998 AD for (I) the optimal QUARTS model, (II) the optimal GLS model, and (III) the GLS model with residual AR structure and number of principal components matched to the optimal QUARTS model.  
\begin{table}[ht]\centering
\begin{tabular}{|r|c|c|c|c|}
\multicolumn{5}{c}{Coefficients with Significance $\alpha$} \\
\hline
Model & $\alpha=10\%$ & $\alpha=5\%$ & $\alpha=1\%$ & $\alpha=0.1\%$ \\
\hline
QUARTS & 48 & 43 & 31 & 16 \\
\hline
AR2 PC3 GLS & 29 & 17 & 12 & 2 \\
\hline
AR1 PC9 GLS & 39 & 32 & 17 & 7 \\
\hline
\end{tabular}
\caption{Number of the 79 proxies which fall at various significance levels for the three different models.}
\label{tab79pval}
\end{table}

It seems the QUARTS model substantially outperforms the GLS model in either case in terms of the number of significant predictors across significance levels.  This is to be expected of a more robust approach, but we also would expect the innovation variance to increase from GLS to QUARTS, since this quantity is explicitly minimized by the GLS solution.  The lower overall reconstruction uncertainty for QUARTS reflects the fact that the lower parameter uncertainty more than makes up for the higher innovation variability. To explore why this may be the case, consider the settings under which GLS has well-defined statistical optimality properties.  In particular, GLS is the maximum likelihood estimator when the innovations are \emph{i.i.d.} normal with constant variance.  This is an approximation that was made for all the models in order to be able to use the parametric bootstrap to quantify uncertainty. However, if we perform the Anderson-Darling test for normality on the QUARTS innovations, we obtain a p-value of 2.0\%. Thus the hypothesis of normality at the standard 5\% significance level is rejected, hence justifying a more robust approach such as QUARTS.  

As a check that the normality assumption in the parametric bootstrap did not substantially affect the uncertainty quantification, the same uncertainty quantification process (including correcting for overfitting) was also implemented using the nonparametric bootstrap.  The resulting out-of-sample prediction interval widths were the same to significant digits for the two GLS models already mentioned, and 0.01 \textdegree C smaller for the QUARTS model. Hence the substantial reduction in uncertainty is retained, and in the process, the stability of the model to the specific uncertainty quantification technique employed is also verified.

A natural question to investigate is to see how our method compares to recently popularized Bayesian Hierarchical Models (BHM), such as BARCAST \citep{Tingley2010}.  We attempted to use BARCAST to fit the entire NH for the same 79 proxies used in our analysis, but due to the spatial structure inherent in the BARCAST model, the problem became extremely high dimensional.  As a result, problems with Markov Chain Monte Carlo (MCMC) convergence prevented us from obtaining independent samples from the posterior, and thus any uncertainty analysis was not feasible.  We note also that the BARCAST model assumes all error terms are Gaussian, and also assumes an isotropic parametric spatial covariance structure. Thus we would expect BARCAST to be less robust than QUARTS.

\section{Nonparametric Modelling of Temperature Conditional Distribution}
\subsection{Introduction and Motivation}
As discussed in detail in the introduction, there is much interest in determining the way in which proxies record temperature. To this end forward models aim to understand and specify the conditional distribution $f(P|T)$ (where $P$ denotes a given proxy, and $T$ denotes temperature). We note that one can also study the condition distribution $f(P|T)$ directly or through Bayes theorem. In particular note that
\[ f(P|T) = \frac{f(T|P) f(P)}{f(T)} \]
So the  target conditional distribution $f(P|T)$ can be inferred if the surrogate quantities $f(T|P)$, $f(T)$, and $f(P)$ are available. Hence obtaining $f(T|P)$ is not only useful for directly obtaining paleoclimate reconstructions or hindcasts, it is also potentially useful for  understanding and specifying forward models. In particular, modeling the conditional quantile distribution  $f(T|P)$ yields non-parametric insights into the entire distribution of $f(P|T)$ and is more flexible than using the least squares approach. Such an endeavor is useful in understanding how proxies respond to climate/temperature extremes.

\subsection{Modelling Temperature Distribution Conditioned on Tree Ring Proxies}
We now proceed to apply the methodology from this paper to the task of modelling the conditional distribution of temperature given a particular proxy type. We focus on tree ring proxies because they are by far the most numerous proxy type.  In the M08 dataset, there are 784 NH paleoclimate tree ring proxy records, the latest of which begins in 1761.  To get a fairly complete view of the temperature distribution conditioned on the tree ring proxies, the $0.1^{th}$, $0.25^{th}$, $0.5^{th}$, $0.75^{th}$, and $0.9^{th}$ quantiles were fit, but in general any quantile $\tau \in (0,1)$ can be fit.

\begin{figure}[ht!]\centering
\includegraphics[scale=.65]{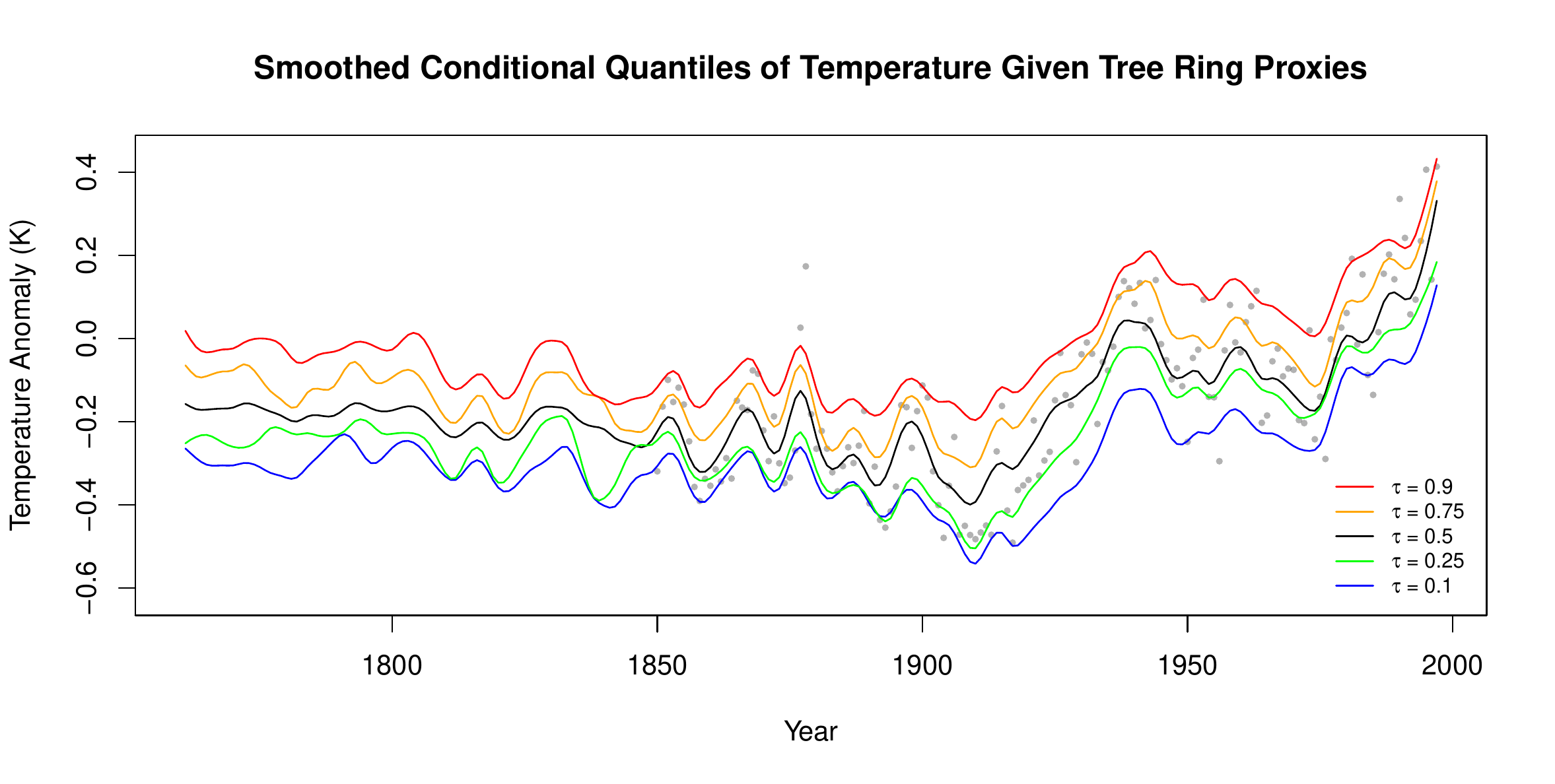}
\caption{Smoothed conditional quantiles of temperature conditioned on tree ring proxies for the period 1761 - 1997.  Instrumental temperatures are plotted as grey points for the period 1850 - 1997.}
\label{TempCondQuantiles}
\end{figure}

The final models for the five quantiles chose 9, 9, 4, 4, and 5 principal components, respectively, with AR(1) residuals.  Figure \ref{TempCondQuantiles} shows the five estimated conditional quantiles over the entire period modelled, with the instrumental data superimposed for the years available.  
The fact that the fitted quantiles generally preserve quantile order is suggestive of our ability to estimate them well.  In other words, they could inform a feasible forward model.

\begin{figure}[ht!]\centering
\includegraphics[scale=.65]{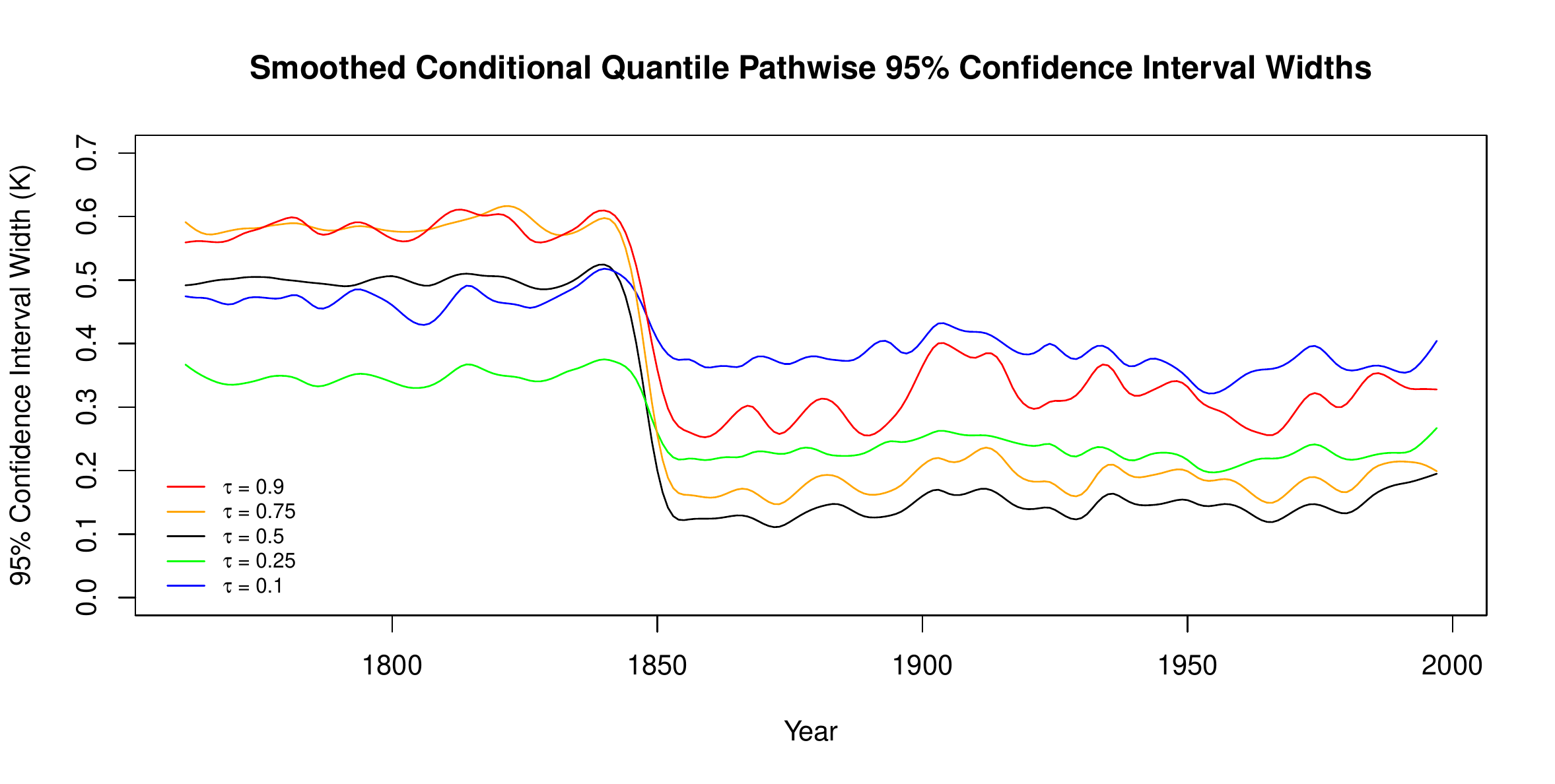}
\caption{Smoothed pathwise 95\% confidence interval widths of conditional quantiles of temperature conditioned on tree ring proxies for the period 1761 - 1997.}
\label{TempCondQuantileCIs}
\end{figure}

Another important component of our analysis is to examine the uncertainty of these conditional quantiles.  Pathwise 95\% confidence interval widths are shown in Figure \ref{TempCondQuantileCIs}.
As expected, during the instrumental period, uncertainty increases as the quantiles diverge from the median.  This is because the model is effectively being fit to fewer points as we move away from the median.  Before the instrumental period, we see some unexpected behavior, with the uncertainty in the lower quantiles ($\tau < 0.5$) being smaller than that in the higher quantiles ($\tau > 0.5$) or even the median.  In other words, the variance of the estimated (conditional quantile) parameter is less in the lower quantiles. Also note the stark difference between before and after 1850 for all of the quantiles.  This occurs because the only uncertainty in the instrumental period comes from the uncertainty in the coefficients, while the out-of-sample uncertainty before 1850 accounts for coefficient uncertainty as well as the randomness in the preceding $y_i$, which are of course not available out-of-sample and on which the conditional distribution depends.

\section{Conclusions}
This paper introduces a novel regression methodology that aims to combine the generalizability to serially correlated residuals of GLS with the robustness and distributional flexibility of QR.  By using the proposed quantile regression with time series errors (QUARTS) method, we are able to create a more robust multiproxy paleoclimate reconstruction than is possible by using traditional conditional-mean-fitting methods.  By extending the bootstrap to the QUARTS framework, we are able to properly quantify uncertainty and show the benefits of this increased robustness.  In addition, by combining the QR and uncertainty quantification we are able to perform a more nuanced analysis of proxy-temperature relationships than has been done previously.

Our reconstruction captures many of the salient features of past reconstructions, and falls well within the uncertainty of the analogous GLS reconstruction and the useful reconstruction of \cite{McShane2011}.  Again the main distinction is that the QR reconstruction has substantially lower uncertainty than any other reconstructions whose uncertainties have been properly statistically accounted for.  

The application of our QUARTS methodology to examining the conditional distribution of temperature given the proxies is chiefly a methodological, as opposed to a scientific, contribution.  We are able to independently generate different quantiles of this distribution and evaluate their uncertainties.  For the example of tree ring proxies, we also show that these conditional quantiles make sense when taken together, and could be used to approximate a full conditional distribution.  It is our hope that proxy specialists such as dendrochronologists, will make use of this to inform and improve our understanding of forward models.

The generality of the methodology introduced in this paper allows it to be easily extended to many other time series regressions outside the earth sciences. However even within the paleoclimate proxy reconstruction context, there is more that can be accomplished.  It has now been nearly a decade and a half since the end of these proxy records, and filling in the intervening years would greatly improve and inform statistical models.  As pointed out in \cite{McShane2011}, one of the shortcomings of paleoclimate reconstructions is their inability to capture the contemporary runup in temperatures, and more points within this runup period would alleviate this problem, at least in part.
\vspace{.5cm}

\emph{Acknowledgements:} Lucas Janson was supported in part by a VPUE grant at Stanford University, and a grant from the Woods Institute for the Environment.  Bala Rajaratnam was supported in part by National Science Foundation under Grant Nos. DMS-CMG 1025465, AGS-1003823, DMS-1106642.  We gratefully acknowledge Michael Tsiang for LaTeX and typesetting assistance, Martin Tingley for discussing BARCAST, and Krista Doersch for help finding a suitable acronym for our methodology. We also thank three anonymous reviewers and an Associate Editor at the Journal of the American Statistical Association for helpful comments that improved the paper.

\bibliography{JASA1_bib}
\bibliographystyle{apalike}

\section*{\Huge{Appendix}}
\appendix
\section{Proof of Lemma 1}
Recall that if a square matrix $\boldsymbol{M} \in \mathbb{R}^{m\times m}$ is positive semidefinite, then any principal submatrix must also be positive semidefinite.  We will show that under the assumptions of Lemma \ref{lemma}, there exists a $2\times 2$ block diagonal submatrix of the full Hessian of the objective function in (\ref{ARQR}) that is not positive semidefinite, hence disproving convexity.

By assumption, let $j \in \{1,\dots,p\}$ satisfy (\ref{ARQRassumpt}).  Define the mixed partial derivative with respect to $\phi_1$ and $\beta_j$ of the objective function in (\ref{ARQR}) as follows:
\[S(\boldsymbol{\phi}, \boldsymbol{\beta}) = \sum_{i=q+1}^{n} \rho'_{\tau}(\phi(B) y_i - \phi(B)\boldsymbol{x}_i^\top \boldsymbol{\beta}) \cdot x_{i-1,j} \]
where $x_{i,j}$ refers to the $j^{th}$ component of the $i^{th}$ observation. Let $\mathcal{A} = \{(\boldsymbol{\phi}, \boldsymbol{\beta}) \in \mathbb{R}^{p+q}$ $s.t.$ $\phi(B) y_i - \phi(B)\boldsymbol{x}_i^\top \boldsymbol{\beta} \neq 0$ $\forall i \in \{q+1,\dots,n\}\}$ be the dense set on which $S(\boldsymbol{\phi}, \boldsymbol{\beta})$ is defined. Recall from (\ref{check}) that $\rho'_{\tau}(.)$, the derivative of the check function, is defined on $\mathbb{R} \backslash \{0\}$ as follows:
\begin{equation*}
\rho'_\tau(x)=\left\{\begin{array}{rl} \tau & $if $x> 0, \\ \tau-1 & $if $x<0. \end{array} \right.
\end{equation*}
It is a simple matter of algebra to show that the matrix of second derivatives with respect to $\phi_1$ and $\beta_j$ of the objective function in (\ref{ARQR}) is equal to 
\[ S(\boldsymbol{\phi}, \boldsymbol{\beta}) \cdot \left(\begin{array}{rl} 0 & 1 \\ 1 & 0 \end{array}\right) \]
on $\mathcal{A}$. The above principal submatrix of the Hessian has eigenvalues $S(\boldsymbol{\phi}, \boldsymbol{\beta})$ and $-S(\boldsymbol{\phi}, \boldsymbol{\beta})$.  By the regularity assumption it is clear that $S(\boldsymbol{\phi}, \boldsymbol{\beta})$ cannot equal zero, and hence one of the eigenvalues of the Hessian above must be negative for all points in $\mathcal{A}$. Therefore, the Hessian of the objective function in (\ref{ARQR}) is not positive semidefinite on the dense set $\mathcal{A}$. We can thus conclude that the objective function is not convex.

\section{Implementation Details for Robust Paleoclimate Reconstruction}
\begin{figure}[ht!]\centering
\includegraphics[scale=.6]{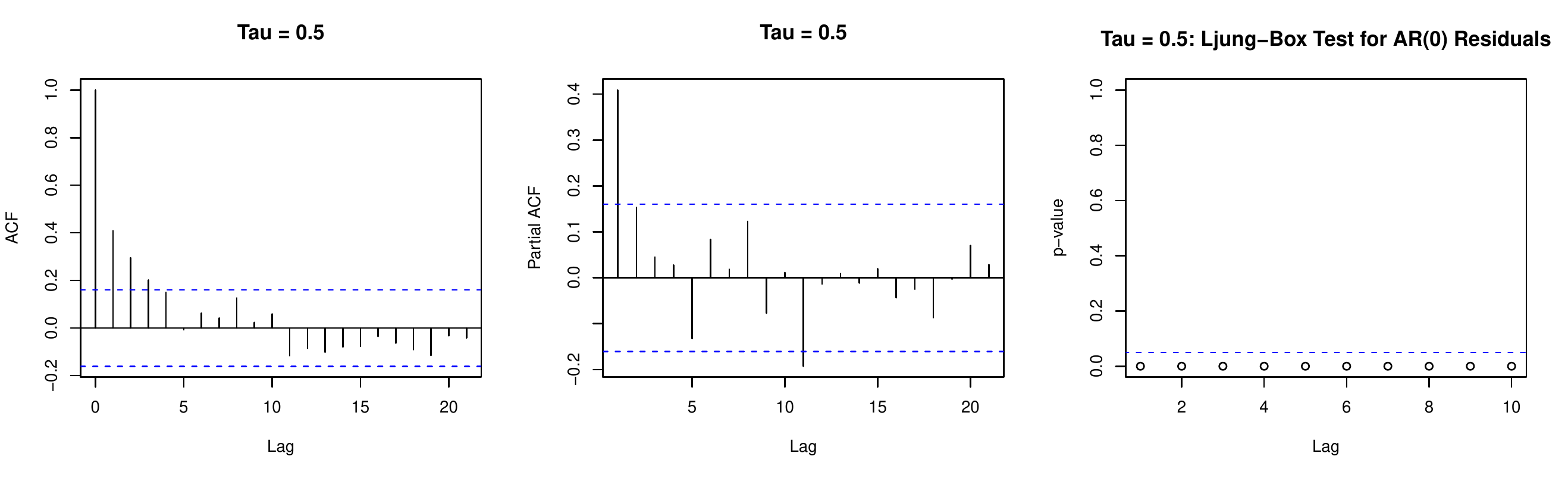}
\caption{Diagnostic plots for assessing AR structure of innovations from a linear model for the conditional median with \emph{i.i.d.} residuals using the 79 oldest proxies. Left: ACF; Center: PACF; Right: Ljung-Box tests.}
\label{79_ARorder0}
\end{figure}

\begin{figure}[ht!]\centering
\includegraphics[scale=.39]{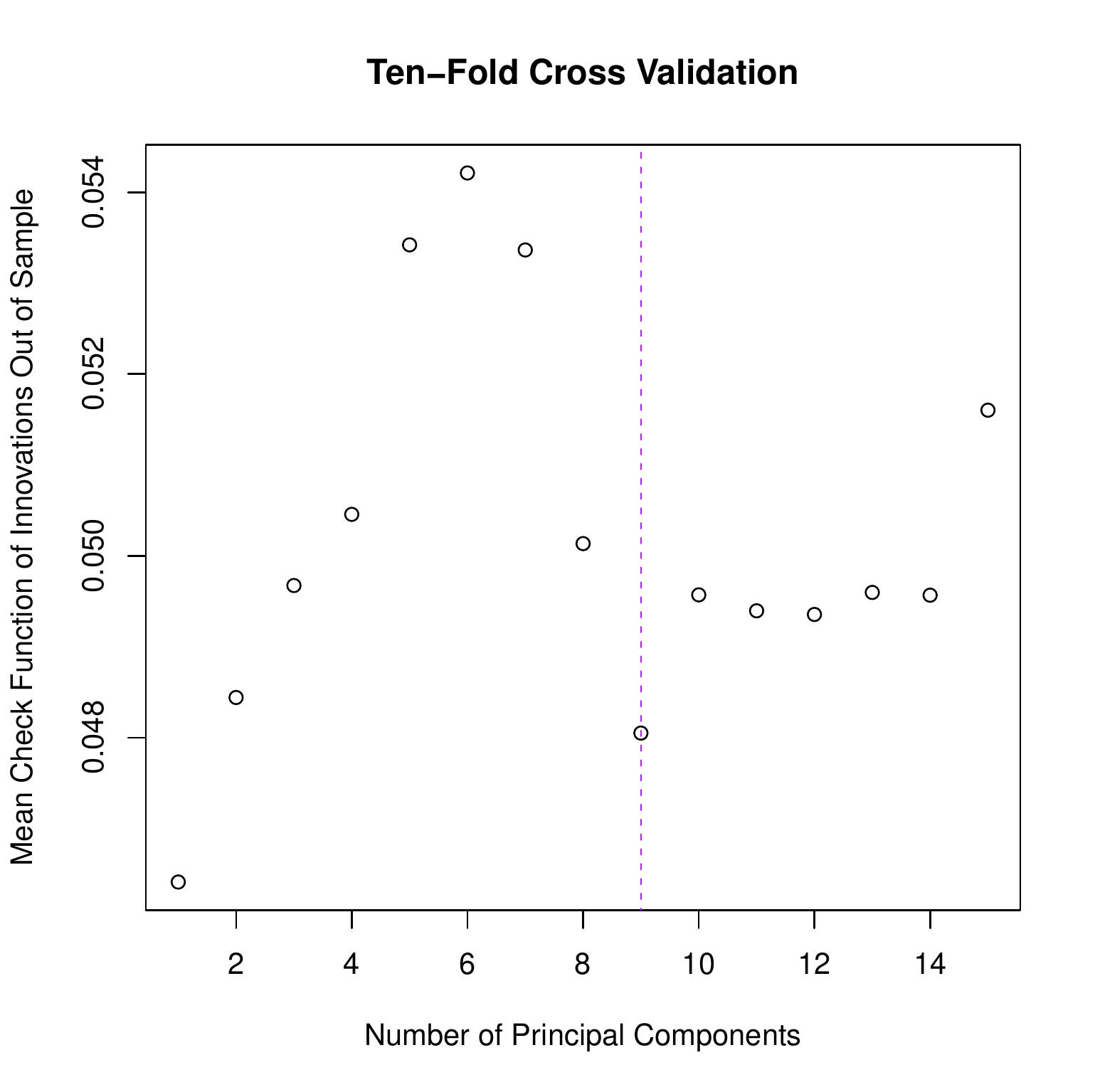}
\caption{Cross validation plot for choosing the number of principal components for fitting the conditional median with AR(1) residuals using the 79 oldest proxies.}
\label{79_choosePCsAR1}
\end{figure}

\begin{figure}[ht!]\centering
\includegraphics[scale=.6]{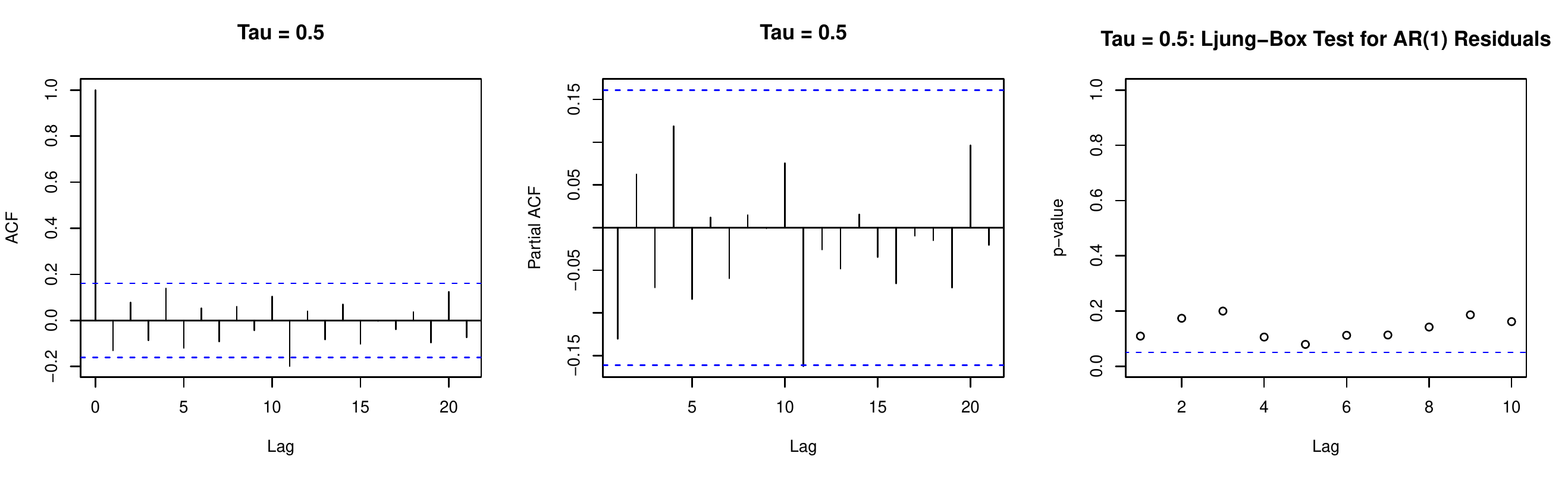}
\caption{Diagnostic plots for assessing AR structure of innovations from a linear model for the conditional median with AR(1) residuals using the 79 oldest proxies. Left: ACF; Center: PACF; Right: Ljung-Box tests.}
\label{79_ARorder1}
\end{figure}

\begin{figure}[ht!]\centering
\includegraphics[scale=.35]{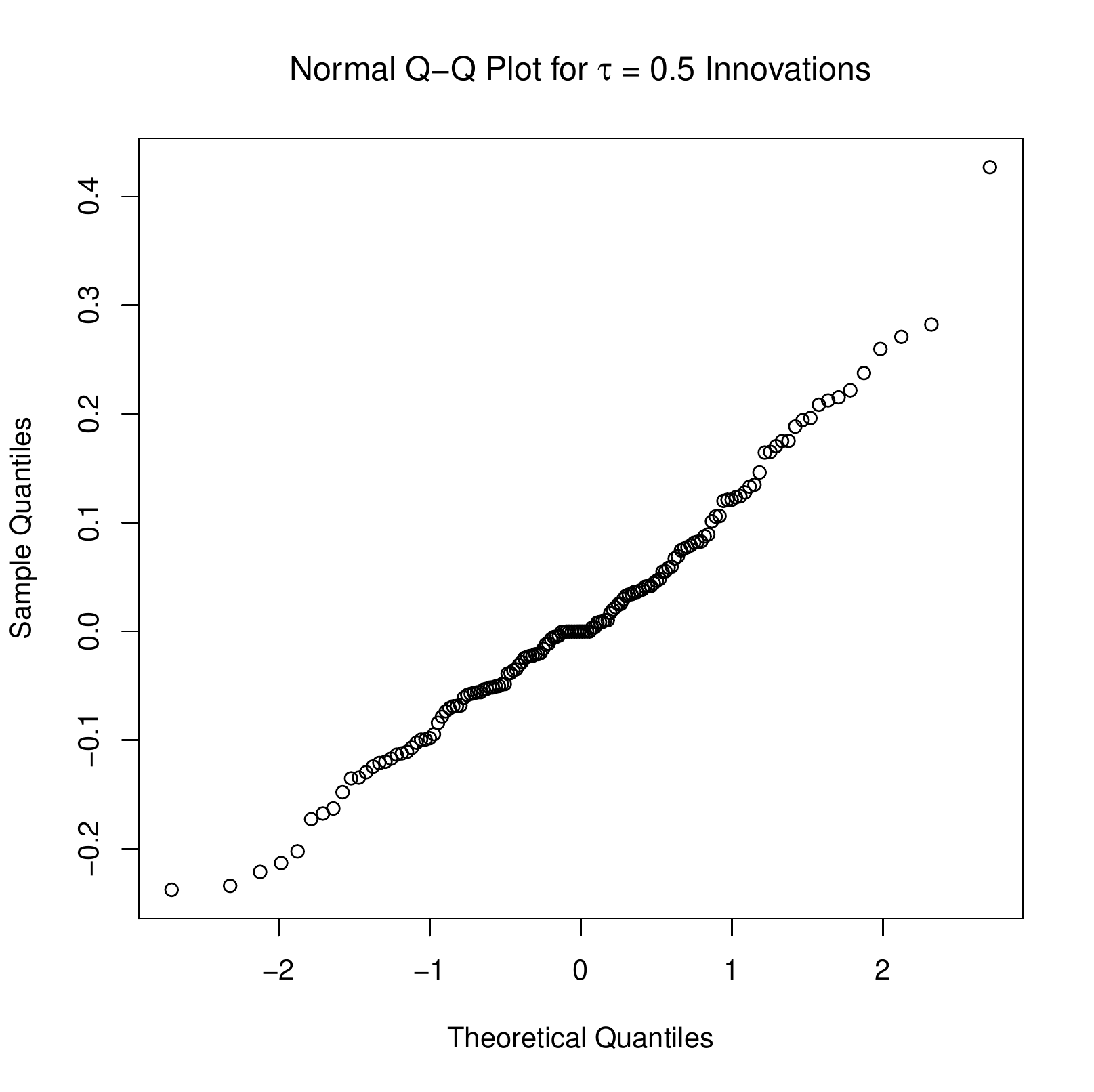}
\caption{Normal quantile-quantile plot for the innovations of the median fit, under the AR(1) assumption.}
\label{79_normalQQ}
\end{figure}

The RARLD algorithm is used to first find a residual AR order.  In particular, an AR(0) model is considered for the residuals: cross-validation chooses 13 principal components.  Diagnostic plots for the innovations from fitting this PCQR model are shown in Figure \ref{79_ARorder0}. We see that AR(0) residuals are rejected on all counts, and hence by the RARLD algorithm an AR(1) model is fitted.  
The AR(1) residual model chooses 9 principal components\footnote{Recall from Section 3.2 that the model is confined to have at least three principal components.}, see Figure \ref{79_choosePCsAR1}, and does not reject the \emph{i.i.d.} innovation hypothesis in any of the three plots shown in Figure \ref{79_ARorder1}.
The final model can now be fitted, but in order to construct pathwise confidence intervals, a parametric distribution for the innovations needs to be specified.  A normal distribution appears to be adequate, as shown in the normal quantile-quantile plot in Figure \ref{79_normalQQ}. A correction however needs to be made for overfitting the variance.  
Table \ref{79_innovParams}  gives the $\hat{\mu}$ and $\hat{\sigma}$ used for the bootstrapping, and the original $\hat{\sigma}_{naive}$ of the empirical innovations before correcting for overfitting.  As expected, the overfitting-corrected standard deviation is larger (by about 17\%) than its na\"{i}ve counterpart.
\begin{table}[ht]\centering
\begin{tabular}{|c|c|c|c|}
\multicolumn{4}{c}{Innovations Parameters} \\
\hline
$\mu$ (\textdegree C) & $\hat{\sigma}_{naive}$ (\textdegree C) & $\hat{\sigma}$ (\textdegree C) & \% Increase\\
\hline
0.012 & 0.111 & 0.130 & 17.1\%\\
\hline
\end{tabular}
\caption{Parameters used in the Gaussian approximation of the bootstrap innovation distribution for the median model using the 79 oldest proxies. The \% Increase measures the inflation between the overfitted $\hat{\sigma}_{naive}$ and the corrected $\hat{\sigma}$.}
\label{79_innovParams}
\end{table}

\end{document}